\documentclass[showpacs,superscriptaddress,twocolumn,aps,reprint]{revtex4-1}
\usepackage{color}
\usepackage{amsfonts,amssymb,amsmath,mathrsfs}
\usepackage{amsbsy}
\usepackage{graphicx}
\usepackage{hyperref}
\usepackage[caption=false]{subfig}

\begin{document}

\title{Stability of skyrmion lattices and symmetries of quasi-two-dimensional chiral magnets}

\author{Utkan G\"ung\"ord\"u}
\email{ugungordu@unl.edu}
\affiliation{Department of Physics and Astronomy and Nebraska Center for Materials and Nanoscience, University of Nebraska, Lincoln, Nebraska 68588, USA}

\author{Rabindra Nepal}
\affiliation{Department of Physics and Astronomy and Nebraska Center for Materials and Nanoscience, University of Nebraska, Lincoln, Nebraska 68588, USA}

\author{Oleg A. Tretiakov}
\affiliation{Institute for Materials Research, Tohoku University, Sendai 980-8577, Japan}
\affiliation{School of Natural Sciences, Far Eastern Federal University, Vladivostok 690950, Russia}

\author{Kirill Belashchenko}
\affiliation{Department of Physics and Astronomy and Nebraska Center for Materials and Nanoscience, University of Nebraska, Lincoln, Nebraska 68588, USA}

\author{Alexey A. Kovalev}
\affiliation{Department of Physics and Astronomy and Nebraska Center for Materials and Nanoscience, University of Nebraska, Lincoln, Nebraska 68588, USA}

%\date{\today}

\begin{abstract}
Recently, there has been substantial interest in realizations of
skyrmions, in particular in quasi-2D systems due to increased stability
resulting from reduced dimensionality. A stable skyrmion, representing
the smallest realizable magnetic texture, could be an ideal element
for ultra-dense magnetic memories. Here, we use the most general form
of the quasi-2D free energy with Dzyaloshinskii-Moriya interactions
constructed from general symmetry considerations reflecting the
underlying system. We predict that skyrmion phase is robust and it is
present even when the system lacks the in-plane rotational symmetry.
In fact, the lowered symmetry leads to increased stability of
vortex-antivortex lattices with four-fold symmetry and in-plane
spirals, in some instances even in the absence of an external magnetic
field. Our results relate different hexagonal and square cell phases to
the symmetries of materials used for realizations of skyrmions. This
will give clear directions for experimental realizations of hexagonal
and square cell phases, and will allow engineering of skyrmions with
unusual properties.
We also predict striking differences in gyro-dynamics induced by spin
currents for isolated skyrmions and for crystals where spin currents
can be induced by charge carriers or by thermal magnons. We find that
under certain conditions, isolated skyrmions can move along the
current without a side motion which can have implications for
realizations of magnetic memories.
\end{abstract}

\maketitle

\section{Introduction}

Skyrmions are topological structures corresponding to highly stable particle-like excitations. Although skyrmions were first invented as a model for baryons \cite{Skyrme1962}, it has been found that their analogs can be realized in condensed-matter systems such as chiral magnets. Existence of magnetic skyrmion lattices has been predicted theoretically \cite{Bogdanov1994} and confirmed experimentally  \cite{Muhlbauer2009a}. Magnetic skyrmions in chiral magnets, which are the main subject of interest in this paper, have received a lot of interest recently \cite{Rossler2006,Yi2009,Muhlbauer2009a,Yu2010,Rybakov2013,Heinze2011,Kiselev2011,Seki2012a,
Ambrose2013,Janoschek2010,Buhrandt2013,Li2014,Kovalev2014a,Kovalev2014b,Han2010,
Banerjee2014,Lin2015,Buttner2015,Ler2010,Tokunaga2015,
Lin2014,Kiselev2011,Wilson2014,Iwasaki2013, Barker2015, Fert2013,Jonietz2011,Schulz2012a}.

One of the most attractive features of skyrmions is their dynamics \cite{Fert2013}. For domain walls in ferromagnets, the threshold current density for current-driven motion is $\sim 10^9-10^{12} \text{A m}^{-2}$ whereas for skyrmions this threshold is $\sim 10^5-10^{6} \text{A m}^{-2}$ in the slow-speed regime \cite{Nagaosa2013}, which may lead to development of low-power and ultra-dense magnetic memories \cite{Tomasello2014,Koshibae2015}. Another attractive feature of skyrmions is the robustness of their motion: a shape-deformation and a Magnus-like force in their dynamics allow skyrmions to avoid impurities and lead to a very robust current-velocity relation \cite{Iwasaki2013}. On the other hand, in magnetic insulators skyrmions can be driven by magnon currents induced by temperature gradients \cite{Kong2013,Lin2014,Mochizuki2014,Kovalev2014a,Kovalev2014b}.
The interlocking of the local magnetization direction and the spin of the conduction electrons in chiral magnets can lead to various transport phenomena such as the topological Hall effect \cite{Neubauer2009,Nagaosa2010,Zang2011} and non-Fermi liquid behavior \cite{Pfleiderer2001,Doiron-Leyraud2003,Watanabe2014}.

In chiral magnets such as MnSi, FeGe, and  Cu$_2$OSeO$_3$ the microscopic spin-orbit coupling (SOC) breaks the inversion symmetry. This chiral interaction prefers twisted magnetic structures rather than uniform magnetization, and eventually can lead to creation of spirals and skyrmions. The SOC manifests itself as the Dzyaloshinskii-Moriya (DM) interaction in the free energy \cite{Dzyaloshinsky1958,Moriya1960,Bak1980,Fert2013}. Symmetries of these magnets determine the nature of the DM interaction and magnetic textures that form within the magnet. For example, broken bulk inversion in noncentrosymmetric materials results in a highly symmetric DM interaction and vortex-like skyrmions. The stability of such skyrmions increases for thin film structures \cite{Yu2011,Butenko2010}, hinting that studies of two-dimensional (2D) systems should be of particular importance. In addition, inversion is naturally broken in 2D systems interfacing between different materials. Recent examples of such systems include magnetic monolayers \cite{Bode2007,Romming2013} as well as magnetic thin films \cite{Dupe2014,Moreau-Luchaire2015} deposited on non-magnetic metals with strong SOC. Magnetic systems at oxide interfaces \cite{Lee2013,Moetakef2012} also reveal large SOC and DM interactions --- ingredients that result in formation of skyrmions \cite{Banerjee2013}. A magnetically doped thin layer on a surface of a topological insulator could be yet another promising system for realizations of  magnetic systems with strong DM interactions \cite{Fan2014,Tserkovnyak2015}. 

In this paper, we study the effect of the structural and bulk asymmetries on the skyrmion (SkX) and square cell (vortex-antivortex) (SC) crystals. Taking the most general form of the free energy with DM interactions, we classify 2D chiral magnets. We discuss possible realizations via appropriate structural asymmetry where microscopically this leads to the appearance of SOC, e.g., of Rashba and/or Dresselhaus type. We calculate the phase diagram for different configurations of DM interactions and find that SkX and SC phases are robust and are present even when the system is lacking the in-plane rotational symmetry --- the case not considered in previous studies \cite{Banerjee2014,Lin2015}.  Magnetization dynamics simulations reveal that skyrmions and vortices flexibly deform and adapt to lowered symmetries, resulting in configurations with unusual shapes. However, we also find that SkX region of the phase diagram gradually shrinks as the asymmetries become stronger.  On the other hand, the lowered symmetry leads to increased stability of the vortex-antivortex SC lattice with four-fold symmetry and the in-plane spiral, in some instances even in the absence of an external magnetic field. In chiral magnets with only reflection symmetry, we find an extremely stable in-plane spiral phase.  

We also address the spin-current-induced dynamics of isolated skyrmions and skyrmion crystals which is of interest due to potential applications of skyrmions in magnetic memory devices \cite{Tomasello2014,Koshibae2015}. Using Thiele's approach, we obtain a general velocity-current relation for skyrmions. Our results show that dissipative corrections can influence the direction of the transverse motion of isolated skyrmions, leading to strikingly different results for isolated skyrmions and skyrmion lattices. Our results apply to spin currents induced by charge carriers as well as to spin currents induced by magnon flows. The latter case is addressed in more detail in Appendix \ref{sec:LLG} where we derive the Landau-Lifshitz-Gilbert (LLG) equation with magnonic torques starting from the stochastic LLG equation. Finally, in Appendix \ref{sec:microscopic} we also show how DM interactions can arise from spin-orbit interactions in microscopic models corresponding to magnets with both localized (relevant to oxide interfaces \cite{Lee2013,Moetakef2012}) and itinerant (relevant to thin magnetic films \cite{Dupe2014,Moreau-Luchaire2015}) spins.

\section{Model}
Our system of interest is a 2D chiral magnet with crystalline anisotropies and in the presence of an external magnetic field. The continuum free energy of such system contains the chiral term known as Dzyaloshinskii-Moriya (DM) interaction \cite{Dzyaloshinsky1958,Moriya1960,Bak1980} which is responsible for broken inversion symmetry. Its origins can be traced back to relativistic spin-orbit coupling (SOC) \cite{Nagaosa2013,Kim2013a,Tserkovnyak2014} which provides a link between structural and magnetic chiralities in the system (see Appendix \ref{sec:chiral} for further discussion).
We phenomenologically introduce the continuum free energy density of the 2D chiral magnet as
\begin{align}
\mathcal F_0 =\sum_{\mu = x,y} \frac{J}{2} (\partial_\mu \boldsymbol n)^2 + (\hat D \boldsymbol e_\mu) \cdot (\boldsymbol n \times \partial_\mu \boldsymbol n),
\end{align}
where $\boldsymbol n$ is a unit vector along local spin density, $\boldsymbol e_\mu$ is a unit vector along the $\mu$-axis in the positive direction, $J > 0$ is the exchange interaction constant, and $\hat D$ is a rank-2 tensor describing the DM interaction whose form depends on the structural and bulk symmetries of the system.
Note that asymmetric spin-orbit interactions can be introduced in a similar way for semiconductor quantum wells \cite{Ganichev2014}.

 For convenience, we work with dimensionless free energy density obtained by the transformation 
 $x \to x/(J/D)$, $y \to y/(J/D)$ where $D = ||\hat D|| > 0$ is the overall strength of the DM interaction and $J/D$ is the typical length scale of magnetic structures such as skyrmions and spirals \cite{Tretiakov2010}. Expressed using the new units of length, the free energy density becomes
\begin{align}
\mathcal F_0 = \sum_{\mu = x,y} \frac{1}{2} (\partial_\mu \boldsymbol n)^2 + (\hat D \boldsymbol e_\mu/D) \cdot (\boldsymbol n \times \partial_\mu \boldsymbol n)
\label{eq:F0}
\end{align}
in units of $J$.
The following discussions in this paper are based on this free energy density with additional Zeeman energy $\boldsymbol H \cdot \boldsymbol n$ due to the external magnetic field and uniaxial anisotropy energy $A_s n_z^2$. Expressed in the new units, the total free energy density is given by
\begin{align}
\mathcal F = \mathcal F_0 + \frac{HJ}{D^2} n_z + \frac{A_s J}{D^2} n_z^2 ,
\label{eq:free-energy}
\end{align}
where $\mathcal F_0$ is given by Eq.~(\ref{eq:F0}), $H=g \mu_B H_a$, $H_a$ is the strength of the applied external magnetic field along the $z$-axis, $g>0$ is the $g$-factor, $\mu_B$ is the Bohr magneton, and $A_s$ is the strength of the anisotropy. Adding moderate strength anisotropy that is not of easy-plane/easy-axis type in Eq.~(\ref{eq:free-energy}) does not change our results qualitatively.

We now investigate the effects of broken symmetries in a chiral magnet due to structural asymmetries, which manifest themselves in the form of $\hat D$ tensor. To this end, let us discuss the correspondence between the structural asymmetries in a system and the form of the free energy density given by Eq.~(\ref{eq:free-energy}).
We first note that in the case of a 2D magnet given that $\partial_z = 0$ (cf. Eq.~(\ref{eq:F0})) the rightmost column of $\hat D$ is unimportant in the sense that it does not contribute to the free energy.
The symmetries of the DM magnet can be classified based on the DM tensor $\hat D$, which can be written as a sum of a symmetric and an antisymmetric tensor as $\hat D = \hat D^\text{sym} + \boldsymbol{D}^\text{asym} \times$. 
The off-diagonal tensor components $[\hat D]_{zx}$ and $[\hat D]_{zy}$ can come from both symmetric and antisymmetric parts of $\hat D$, however we can assume that they are due to the antisymmetric part without losing any generality in our classification. Thus assuming $[\hat D^\text{sym}]_{zx} = [\hat D^\text{sym}]_{zy} =0$, the $2 \times 2$ upper-left block of $\hat D^\text{sym}$ can be diagonalized and expressed as $D_0 \openone + D_3 \hat \lambda_3$ (here $\hat \lambda_3 = \text{diag}(1,-1,0)$ and $\openone = \text{diag}(1,1,0)$) by an in-plane O(2) operation (rotation and/or reflection) around the $z$-axis. Since $[\hat D]_{zz}$ does not enter into the free energy, this means $\hat D^\text{sym}$ can be specified using two independent parameters ($D_0, D_3$). The antisymmetric part, however, requires the full set of three parameters $\boldsymbol{D}^\text{asym}=(D_x, D_y, D_z)$.

\begin{figure}[!htb]
\centering
\includegraphics[width=0.65\columnwidth]{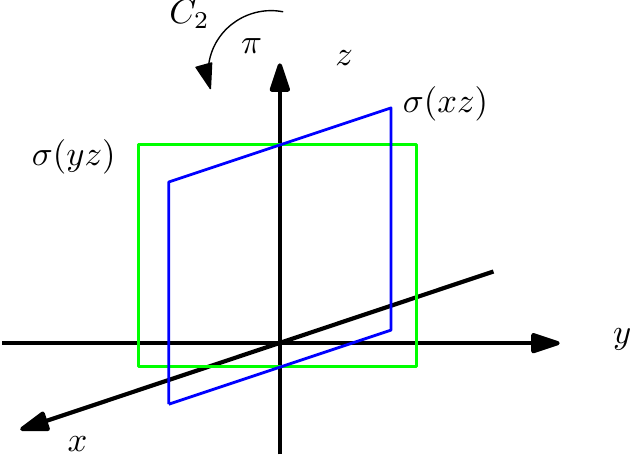}
\caption{(Color online) Nontrivial symmetry operations of the point group $C_{2v}$. $\sigma(yz)$ ($\sigma(xz)$) is a reflection through the $yz$ ($xz$) plane and $C_2$ is a $\pi$ rotation around the $z$-axis.}
\label{fig:digraph}
\end{figure}
Based on this decomposition, symmetries of a 2D DM magnet can be classified into six symmetry classes \footnote{The symmetry classes with $\hat D= \hat J_z, \openone, \hat\lambda_1, \hat\lambda_3$ are related to the 3D classes classified by the Lifshitz invariants $w_1, w_2, w_1', w_2'$, which were described in \cite{Bogdanov1989}}:
\begin{enumerate}
\item Rashba SOC: $\hat D = -D \boldsymbol{z}\times=-D \hat J_z$ with in-plane O(2) symmetry (throughout this paper, $\hat J_\mu$ denote the generators of SO(3), explicitly given in Eq.~(\ref{eq:so3-generators})).
\item Dresselhaus SOC: $\hat D =  -D \hat \lambda_1$ ($\hat \lambda_1$ is given in Eq.~(\ref{eq:reflection}) with $D_{2d}$ symmetry. \footnote{The Hamiltonian (see Appendix \ref{sec:chiral}) for Rashba and Dresselhaus SOC is more often represented as $\alpha (p_x \sigma_y - p_y \sigma_x) + \beta (p_x \sigma_x - p_y \sigma_y)$ than $\alpha (p_x \sigma_y - p_y \sigma_x) + \beta (p_x \sigma_y + p_y \sigma_x)$ in the literature. One can be transformed into another by an in-plane $\pi/4$ rotation. We use the latter form in order to align the mirror symmetry planes with $x$- and $y$-axes, which is compatible with the periodic boundary conditions of our simulations.}
\item In-plane SO(2) symmetry: $\hat D =  -D_0 \openone$ resulting in the DM interaction term $-D \boldsymbol n \cdot (\boldsymbol \nabla \times \boldsymbol n)$ which could be of relevance, e.g., for MnSi \cite{Bak1980,Yi2009}. This case could also include a Rashba contribution $\hat D =  -D_0 \openone -D_z \hat J_z$ without affecting the symmetry class.
\item Combination of Rashba and Dresselhaus SOC, found in noncentrosymmetric systems: $\hat D = -D_R \hat J_z - D_D \hat \lambda_1$ such that $D=\sqrt{D_R^2 + D_D^2}$, corresponding to an interface with $C_{2v}$ symmetry ($C_{2v}$ symmetry is described in Fig.~\ref{fig:digraph}).
\item $C_2$ symmetry: $\hat D = -D_0 \openone - D_1 \hat \lambda_1-D_z \hat J_z$.
\item Reflection symmetry: $\hat D = -D \boldsymbol n_T \times$ where $\boldsymbol n_T$ is a tilted unit vector making a nonzero angle with the $z$-axis. For such a system, the plane of reflection is along the in-plane component of $\boldsymbol n_T$. This case could also include a contribution from the Dresselhaus term when the tilting is along one of the mirror planes of $C_{2v}$ symmetry. 
\end{enumerate}

\begin{figure}[t]
\centering
$Q=1$

\subfloat[$\gamma=0$]{
\includegraphics[width=0.2\columnwidth]{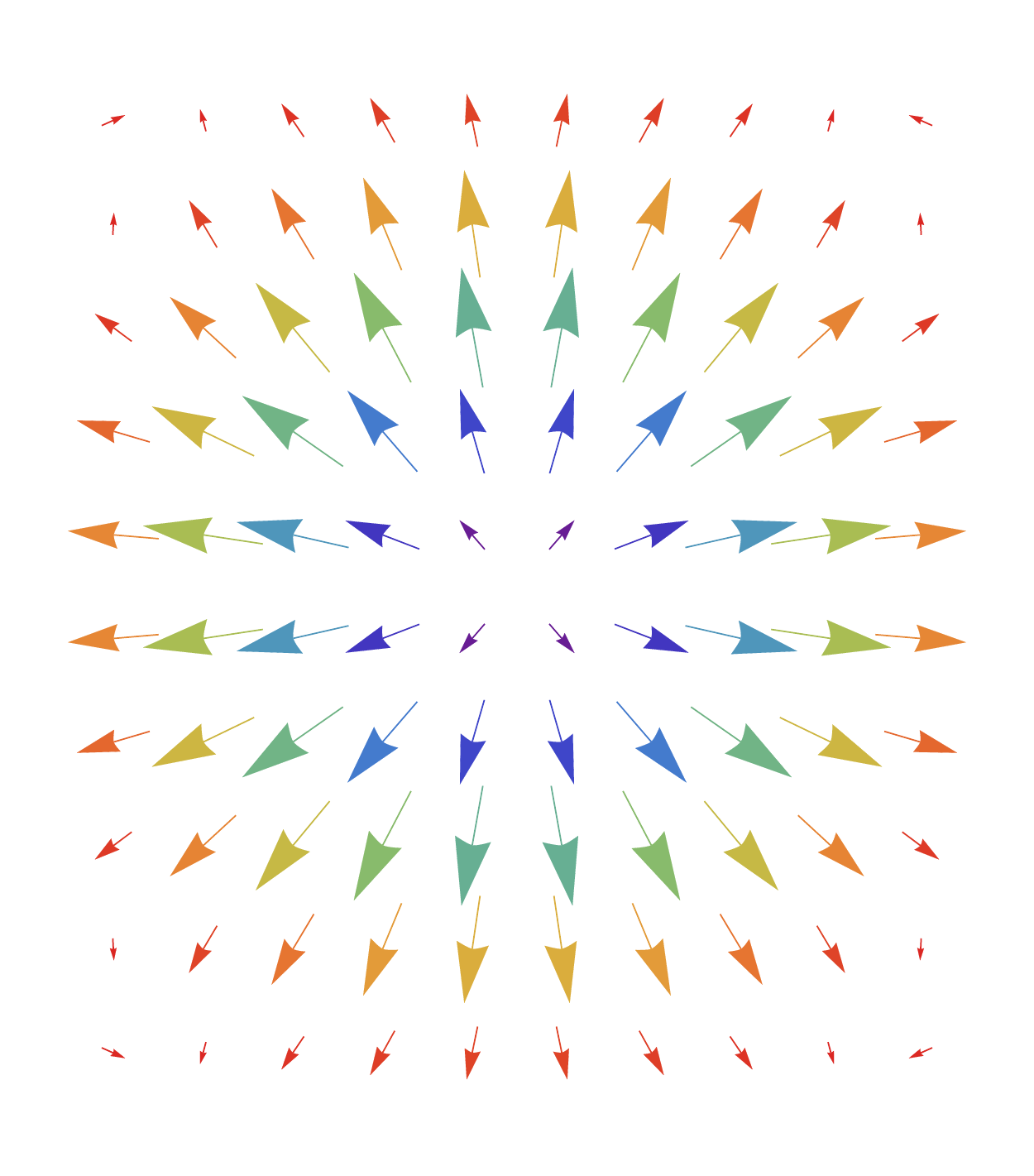}
\label{fig:skyrmion-rashba}
}
\subfloat[$\gamma=\frac{\pi}{2}$]{
\includegraphics[width=0.2\columnwidth]{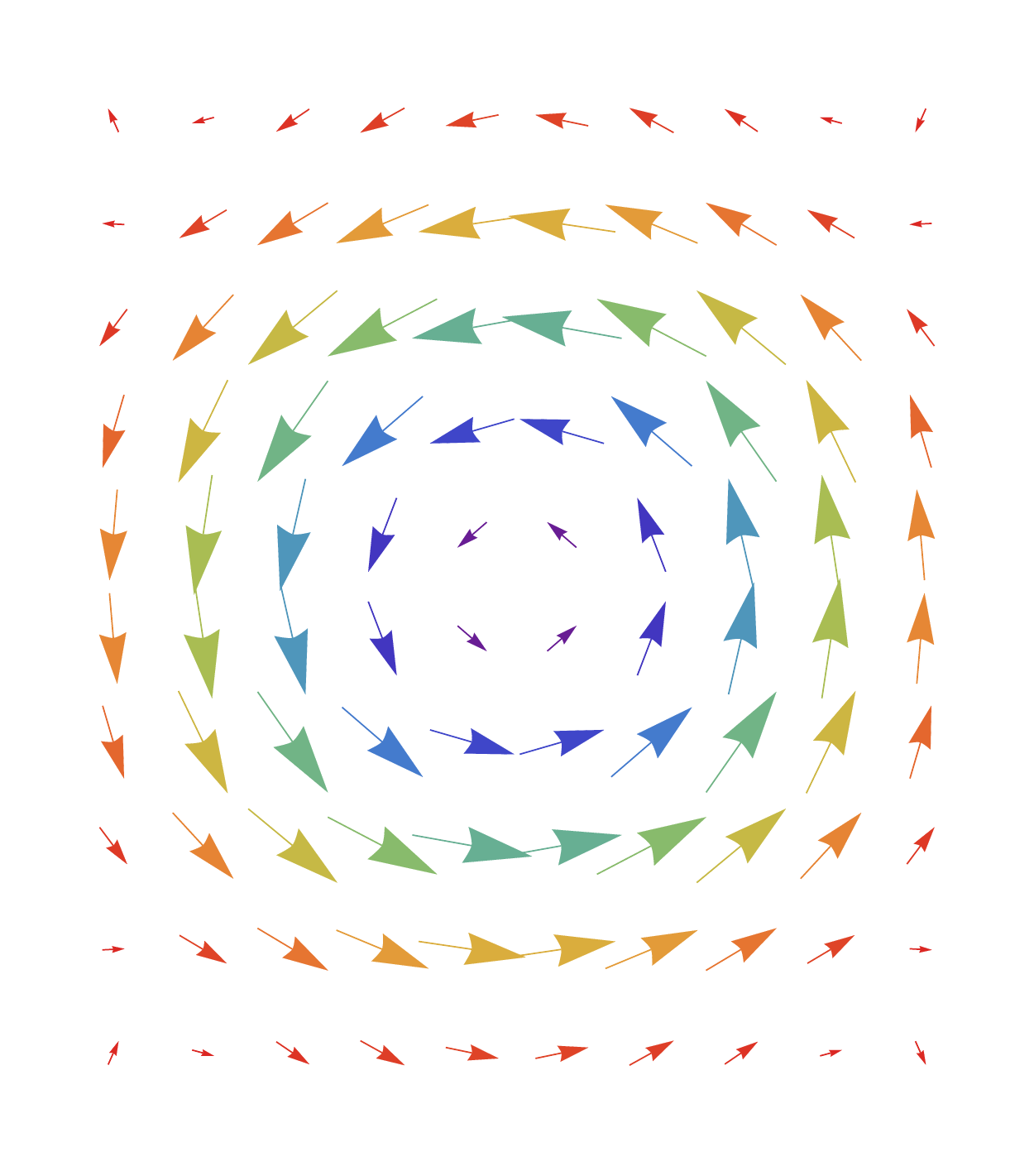}
}
\subfloat[$\gamma=\pi$]{
\includegraphics[width=0.2\columnwidth]{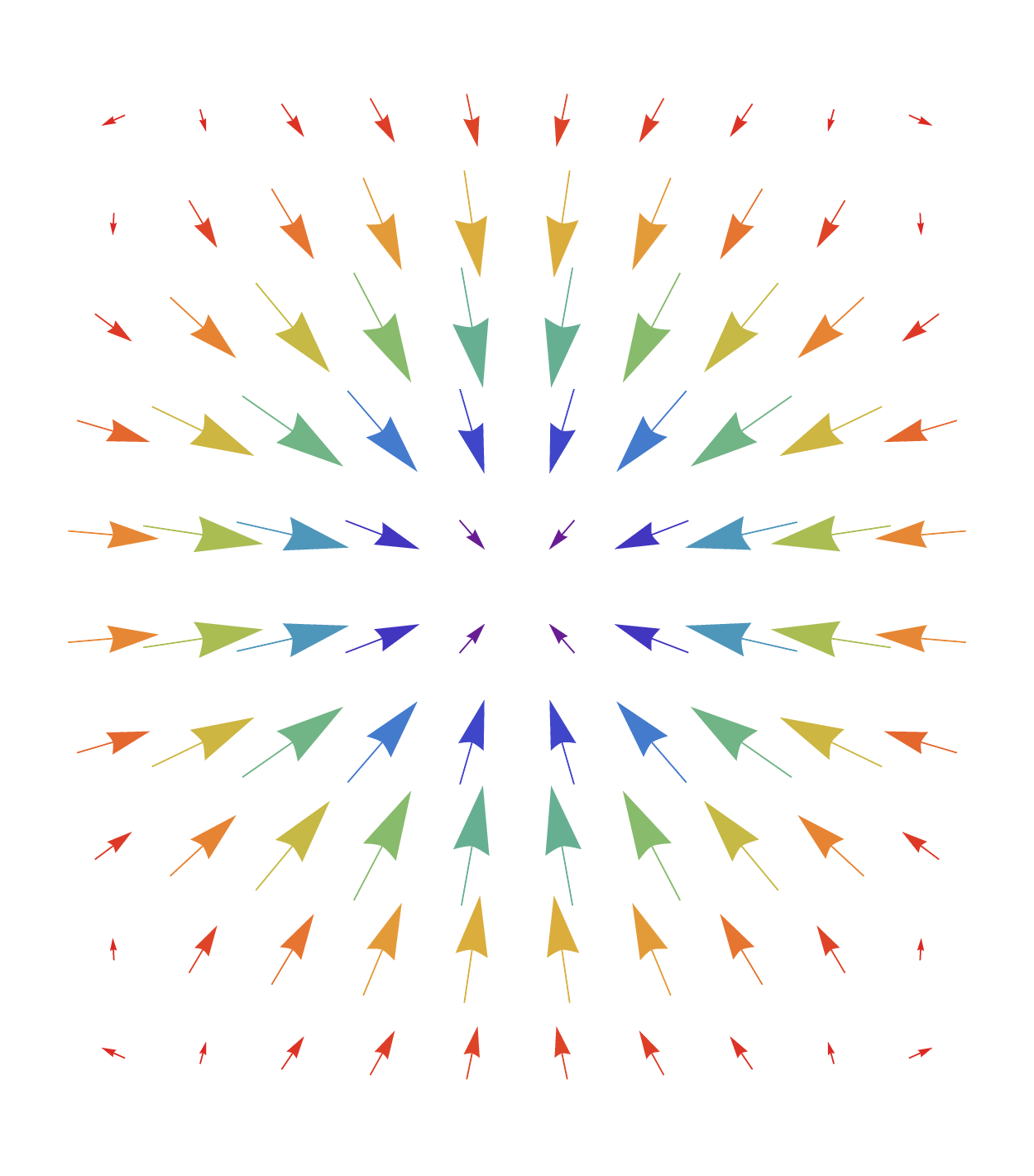}
}
\subfloat[$\gamma=-\frac{\pi}{2}$]{
\includegraphics[width=0.2\columnwidth]{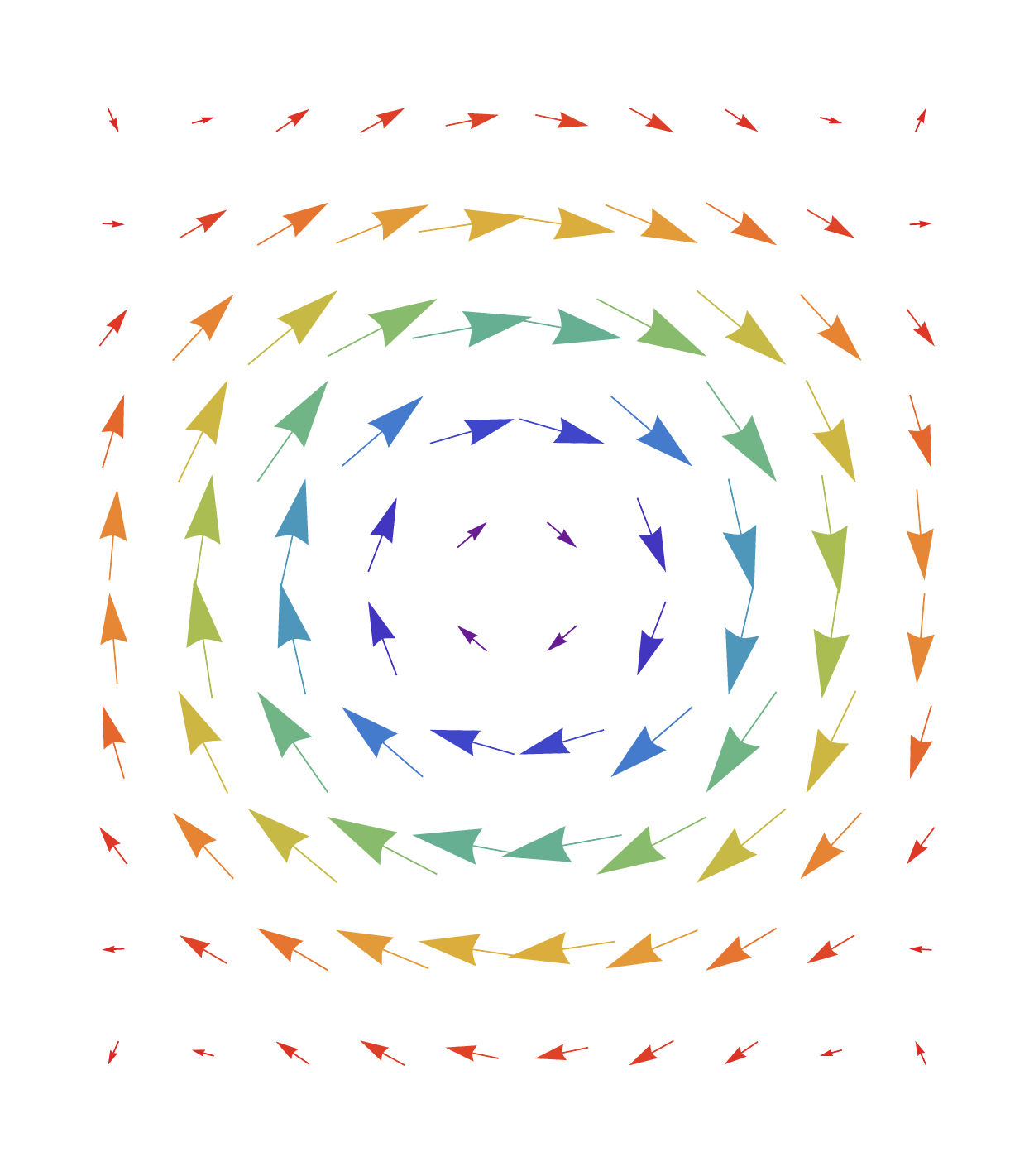}
\label{fig:skyrmion-identity}
}
\vspace{10pt}
$Q=-1$

\subfloat[$\gamma=0$]{
\includegraphics[width=0.2\columnwidth]{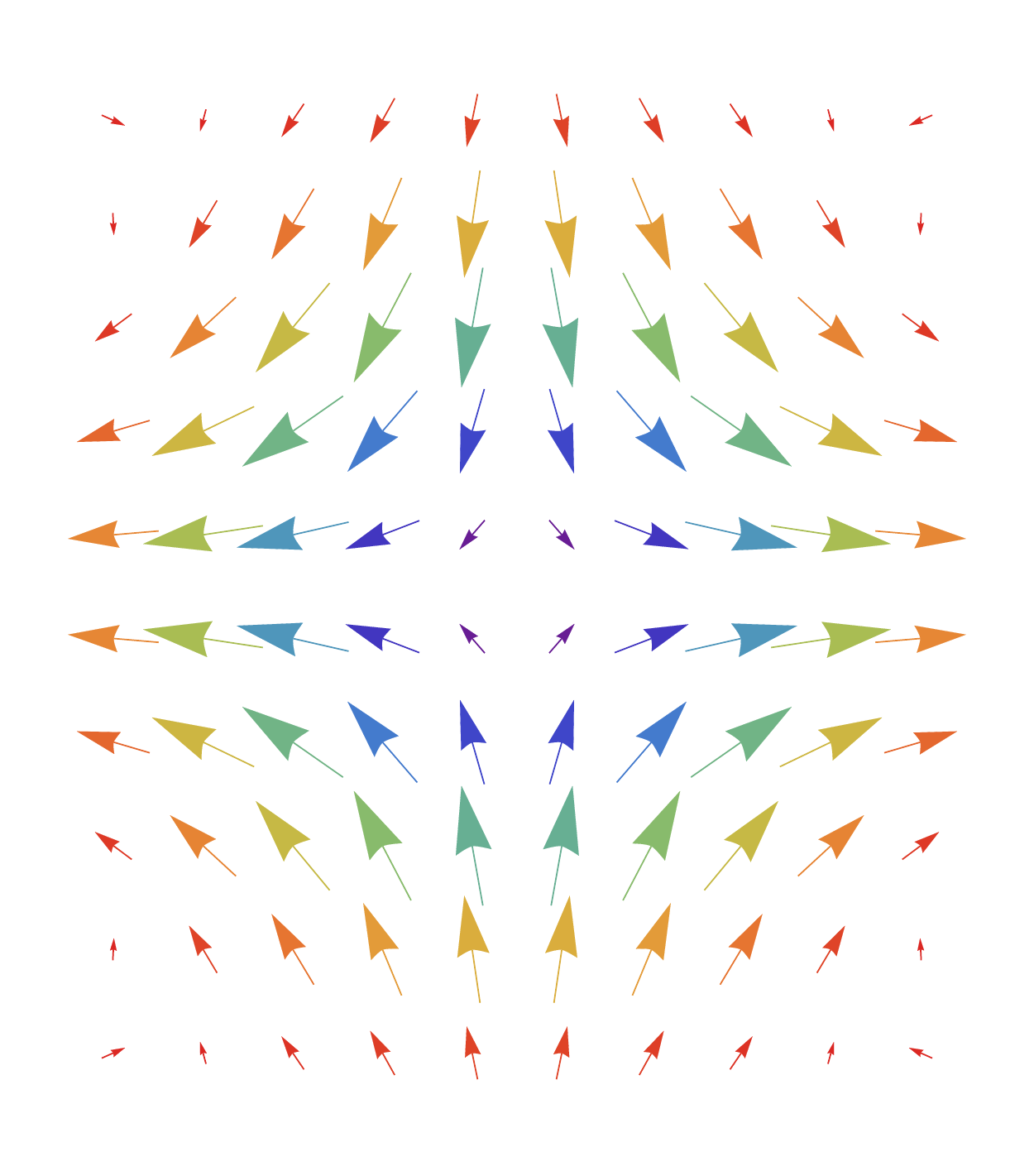}
\label{fig:skyrmion-dresselhaus}
}
\subfloat[$\gamma=\frac{\pi}{2}$]{
\includegraphics[width=0.2\columnwidth]{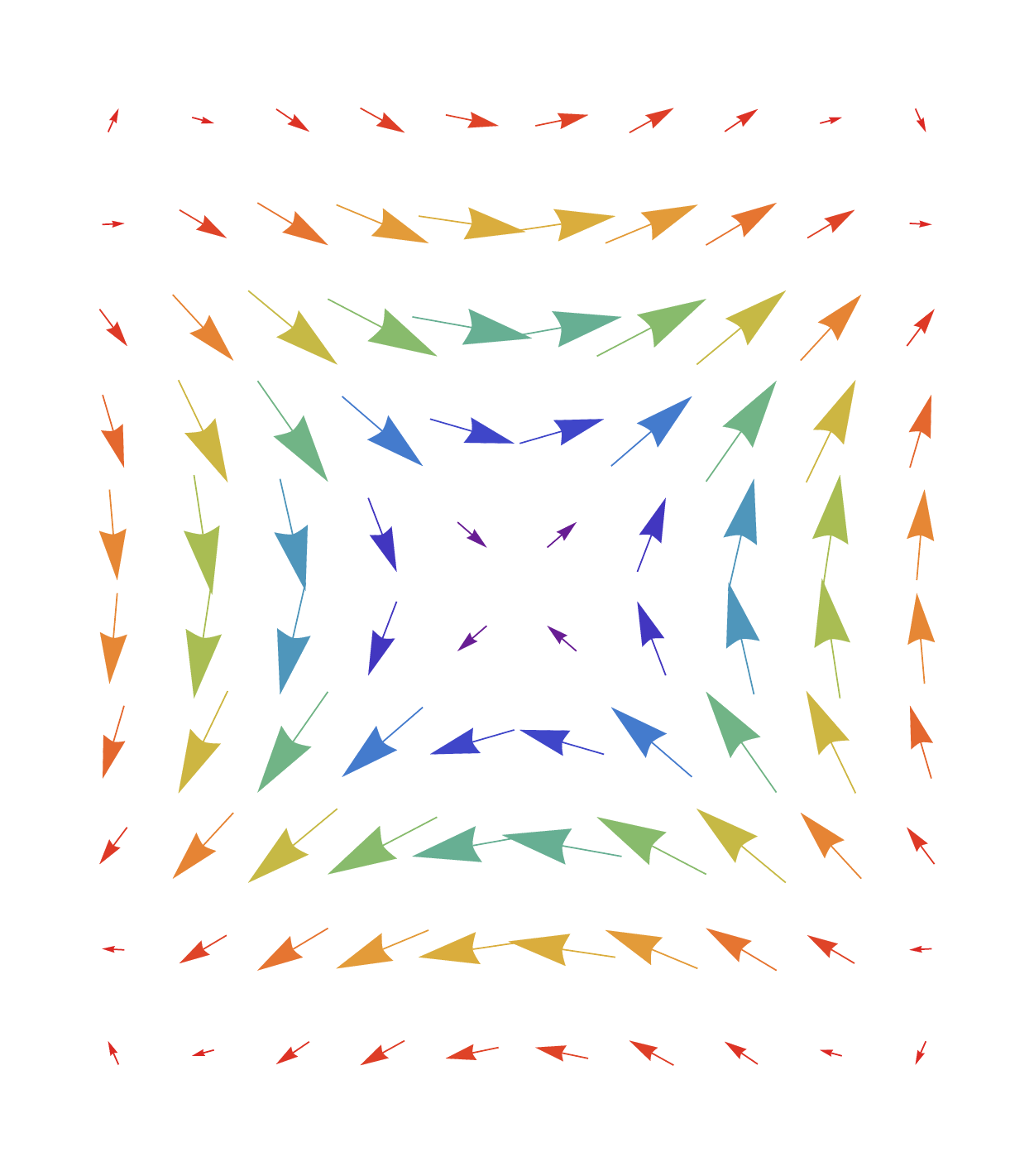}
}
\subfloat[$\gamma=\pi$]{
\includegraphics[width=0.2\columnwidth]{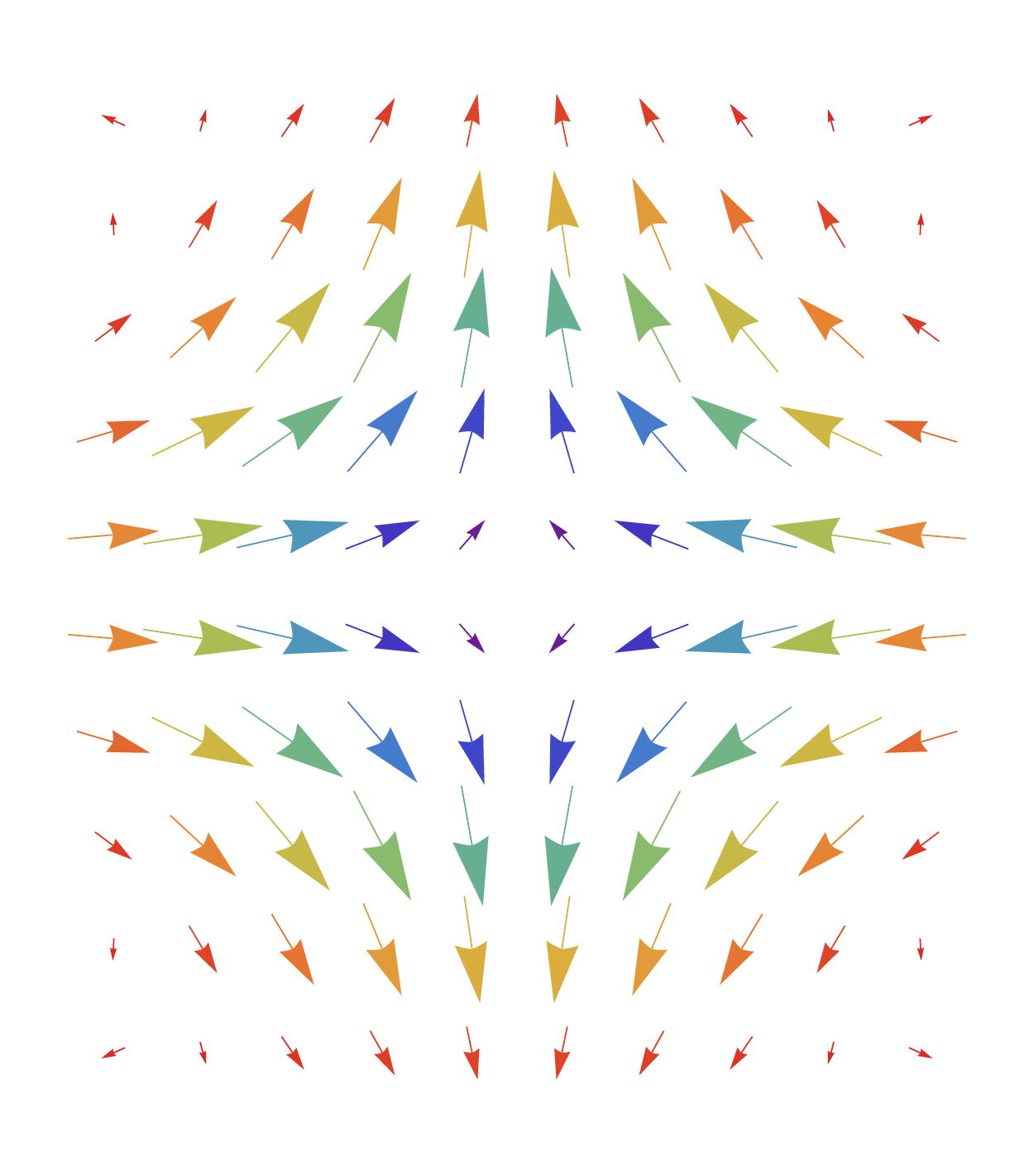}
}
\subfloat[$\gamma=-\frac{\pi}{2}$]{
\includegraphics[width=0.2\columnwidth]{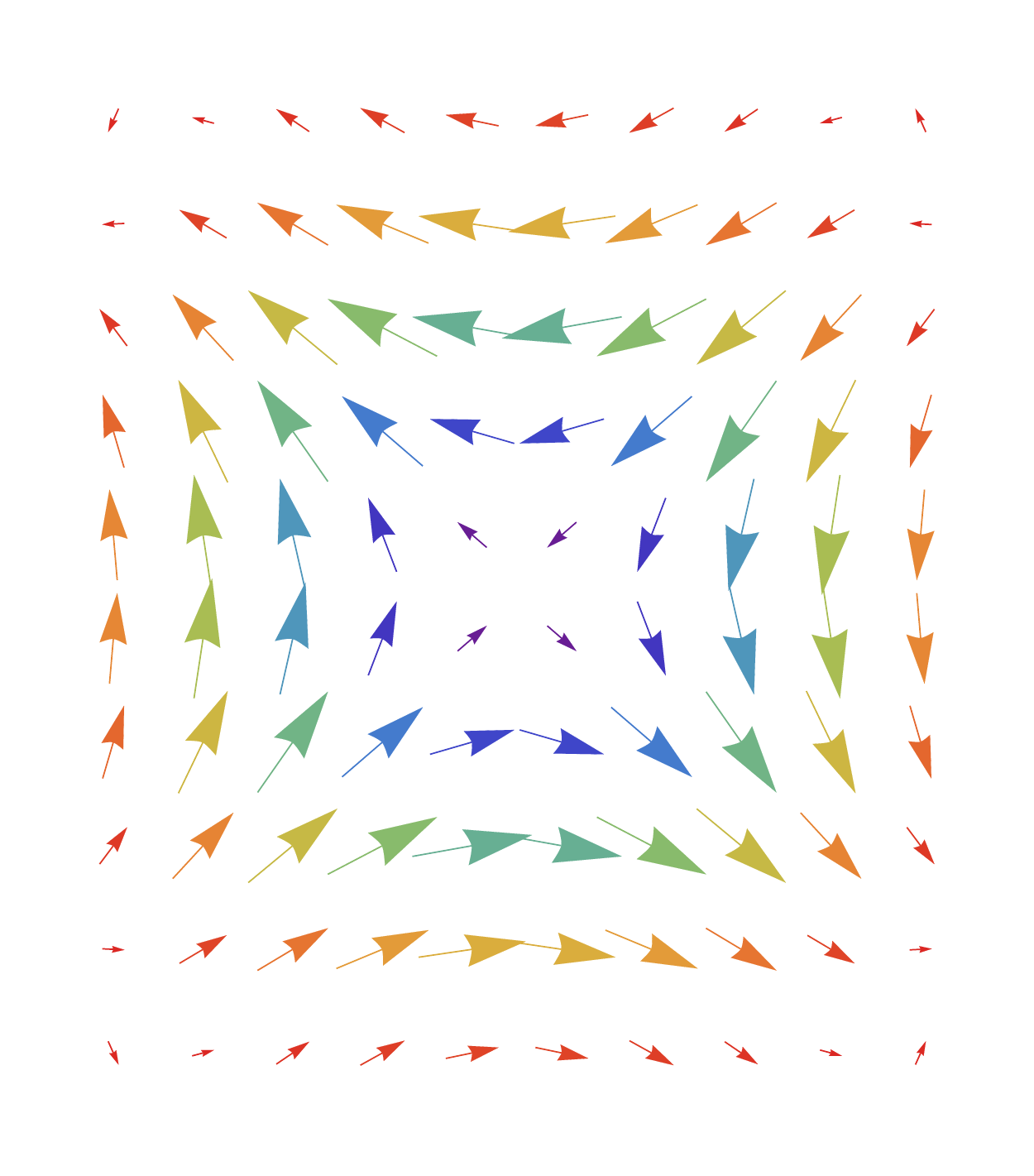}
}
\caption{(Color online) Spin density configuration of skyrmions corresponding to different SOC. The length and direction of the arrows represent the in-plane component of $\boldsymbol n$, and the color indicates $n_z$. At the skyrmion center, $n_z$ is aligned with the external magnetic field $\boldsymbol H = H \boldsymbol e_z$, whereas in the outer region of skyrmion $n_z$ is antialigned with $\boldsymbol H$. The spin configuration $\boldsymbol n_{\hat J_z}$ shown in Fig.~\ref{fig:skyrmion-rashba} represents a skyrmion with topological charge $Q=1$ and helicity $\gamma=0$, which occurs for Rashba SOC given by $\hat D = -D\hat J_z$. Other skyrmions [antiskyrmions] with topological charge $Q=1$ [$Q=-1$] and an arbitrary helicity $\gamma$ can be obtained via the global transformation $\hat R_z(\gamma) \boldsymbol n_{\hat J_z}$; they occur naturally in a system with DM tensor $\hat D = -D\hat R_z(\gamma) \hat J_z$ [$\hat D = D\hat R_z(\gamma) \hat \lambda_3 \hat J_z$)], where $\hat R_z(\gamma)=\exp(\gamma \hat J_z)$ is a rotation around the $z$-axis by an angle $\gamma$ and $\hat \lambda_3=\text{diag}(1,-1,0)$ is an inversion matrix.}
\label{fig:skyrmions}
\end{figure}

Let us turn to the properties of the free energy under global transformations of magnetization. The free energy density of the DM magnet given by  Eq.~(\ref{eq:free-energy}) does not change under the global transformation $\boldsymbol n \to \hat R_z \boldsymbol n$, $\hat D \to \hat R_z \hat D$, where $\hat R_z \in \text{O}(2)$ is a global rotation around/reflection through the $z$-axis (note that this is not a similarity transformation). It follows that the cases $\hat D = -D \boldsymbol z \times$ and $\hat D = -D \openone$ (or their linear combinations) can be mapped onto each other and lead to the same phase diagram \footnote{In a 3D system, the cases $\hat D = -D \openone_3 = -D\text{diag}(1,1,1)$ and $\hat D = -D \hat J_z$ are not equivalent and result in different phase diagrams \cite{Rowland2015}.} where the corresponding mapping is given in Figs.~\ref{fig:skyrmion-rashba} and \ref{fig:skyrmion-identity}. These two cases have been studied in \cite{Banerjee2014} and \cite{Lin2015,Yi2009}, respectively. A global magnetization rotation changes the helicity of the skyrmions by the angle of rotation, whereas a reflection changes the sign of their topological charge (for examples of the equivalent skyrmion configurations and the corresponding DM interactions see Fig.~\ref{fig:skyrmions}). Clearly, the overall sign of the DM term is unimportant in the sense that it does not affect the phase diagram.
% The overall sign of $\hat D$ not of importance as it simply describes a system with opposite chirality.
Another example is given by the equivalence between the DM tensors $\hat{D}=-D_R \hat J_z - D_D \hat \lambda_1$ and $\hat{D}=-D_R \openone - D_D \hat \lambda_3$, which are related by a global $-\pi/2$  rotation of magnetization around the $z$-axis. We can also relate the free energies of systems with Dresselhaus SOC and the Rashba SOC (corresponding skyrmions are shown in Figs.~\ref{fig:skyrmion-dresselhaus} and \ref{fig:skyrmion-rashba}, respectively) using a reflection along the line $x=y$:
\begin{align}
\hat \lambda_3 = \hat R_{x=y} \hat J_z \quad \text{with} \quad \hat R_{x=y} = \hat \lambda_1 =
\begin{pmatrix}
0 & 1 & 0 \\
1 & 0 & 0 \\
0 & 0 & 0
\end{pmatrix}.
\label{eq:reflection}
\end{align}
Here we dropped the $[\hat R_{x=y}]_{zz}$ component of the reflection matrix since it does not play any role in this context. Owing to the transformation rules of the free energy density given by Eq.~(\ref{eq:free-energy}), phase diagrams corresponding to pure Rashba and Dresselhaus SOC are identical. Note that Rashba and Dresselhaus skyrmions will have opposite topological charge.

We finally note that since the rightmost column of the DM tensor $\hat D$ does not affect the free energy, whether $[\hat D]_{zx}$ and $[\hat D]_{zy}$ belong to the antisymmetric part of the tensor or not does not affect the phase diagram either.
 
Thus there are only three distinct symmetry classes we need to consider: the Rashba case, the Rashba combined with Dresselhaus case, and the Rashba with tilting case, i.e. $\hat D = -D \boldsymbol n_T \times$.
We remark that the equivalence relations we just described are global transformations relating the free energies of two distinct systems with different DM tensors, whereas the symmetry classes we enumerated above correspond to the real-space symmetries of the free energy of the system corresponding to the specified DM tensor.

\section{Phase diagrams}
\label{sec:phase-diagrams}
In this section, we present our results for the high- and low-symmetry cases of the spin-orbit interaction. 
We used Monte Carlo (MC) simulated annealing with jackknife resampling to obtain the equilibrium states. To construct the phase diagrams $(A_s, H)$, one can compare the energy of the states obtained from low-temperature ($k_B T = 0.01J$) annealing against the zero temperature ground-state energy of the uniform ferromagnetic state \cite{Lin2015}. However, all the phase transitions we study are of the first order (except for the one between the collinear aligned and tilted FM phases, which is a second order transition). As the first order phase transitions exhibit hysteresis, the annealing does not give clear results near the phase boundaries. Instead, the configuration gets stuck in metastable states, limiting the accuracy in finding the critical values of $A_s$ and $H$, and allowing only a semi-quantitative analysis of the phase diagram \cite{Lin2015}. On the other hand, the advantage of annealing over the variational approach \cite{Banerjee2014} is that ansatz-based minimization may miss certain phases such as the square-lattice phase reported in \cite{Lin2015}. Our approach here is two-fold. We first used MC to determine the phases and spin density configurations, and to obtain a sketch of the phase diagram. Using this information as the starting point, we solved the LLG equation numerically to determine the phase boundaries accurately at zero temperature. 

The spin density configuration $\boldsymbol n(\boldsymbol r_i)$ obtained from MC simulations has been analyzed by inspecting the Bragg peaks in the momentum space
\begin{align}
\boldsymbol n(\boldsymbol k) = \int \frac{d^d \boldsymbol r}{(2 \pi)^d} e^{-i \boldsymbol k \cdot \boldsymbol r} \boldsymbol n(\boldsymbol r),
\end{align}
as well as the topological charge density, which is given by
\begin{align}
\chi(\boldsymbol r) = \frac{1}{4 \pi} [\partial_x \boldsymbol n(\boldsymbol r) \times \partial_y \boldsymbol n(\boldsymbol r)] \cdot \boldsymbol n(\boldsymbol r),
\end{align}
where $\boldsymbol r$ and $\boldsymbol k$ are (in-plane for $d=2$) position and momentum vectors and $d=2$ for a 2D system.

Annealing results reveal the presence of four different phases: ferromagnetic (FM, aligned with $n_z = -1$ and tilted with $n_z > -1$), triangular skyrmion lattice (or skyrmion crystal, SkX), square cell (SC) lattice of vortices-antivortices (with the topological charge that is not an integer or a half-integer), and spiral (SP, denoting both coplanar and in-plane spirals) phases.

Additional phases such as the cone phase  may occur in a 3D chiral ferromagnet with the width $\sim J/D$ or thicker \cite{Rowland2015,Wilson2014}. The presence of quartic term in the free energy near the Curie temperature $T_c$ is also known to influence the phase diagram \cite{Li2014}.

In order to determine the phase boundaries in a more precise manner, we numerically solved the overdamped LLG equation
\begin{align}
s (1+ \alpha \boldsymbol n \times) \dot {\boldsymbol n} = \boldsymbol n \times\boldsymbol H_\text{eff},
\label{eq:overdamped-LLG}
\end{align}
in order to relax the system towards the local minimum. Here $\alpha$ is the Gilbert damping parameter, $s$ is the local spin density, $\boldsymbol{n}$ is a unit vector along the spin density, and $\boldsymbol H_\text{eff} = -\delta_{\boldsymbol n} F$ is the effective magnetic field and $F$ is the free energy.
We used the LLG equation to relax the system into a stable state, starting from SC, SkX and SP configurations that are based on the results from MC:
\begin{align}
\boldsymbol{n}_\text{SkX} &= C[ \boldsymbol n_q(\boldsymbol Q_0) + \boldsymbol n_q(\boldsymbol Q_{2\pi/3}) + \boldsymbol n_q(\boldsymbol Q_{-2\pi/3})], \nonumber \\
\boldsymbol{n}_\text{SC} &= C[ \boldsymbol n_q(\boldsymbol Q_{0}) +  \boldsymbol n_q(\boldsymbol Q_{\pi/2})], \nonumber \\
\boldsymbol{n}_\text{SP} &= \frac{1}{\sqrt 2}[\boldsymbol e_u \cos(\boldsymbol q \cdot \boldsymbol r) + \boldsymbol e_v \sin(\boldsymbol q \cdot \boldsymbol r)],
\label{eq:ansatz}
\end{align}
where
\begin{align}
\boldsymbol n_{q}(\boldsymbol q) &= \cos(\boldsymbol q \cdot \boldsymbol r)\boldsymbol e_z + \frac{1}{2}\sin(\boldsymbol q \cdot \boldsymbol r) \boldsymbol e_z \times \left(\frac{\hat D}{D} \boldsymbol q\right), \nonumber \\
\boldsymbol Q_{\phi} &= (\cos\phi, \sin\phi,0)^T, \nonumber \\
\boldsymbol r &= (x,y,0)^T, \nonumber \\
\boldsymbol e_u \cdot \boldsymbol e_v & = 0,
\end{align}
and $C$ is a (position-dependent) normalization factor ensuring that $|\boldsymbol{n}|=1$. Initial values for the unit vectors $\boldsymbol e_u$ and $\boldsymbol e_v$ were determined by minimization, and the cell size was treated and optimized as a dynamical variable (see Appendix \ref{sec:scaling} for details).
We determined the phase at each $(A_s, H)$-point by comparing the average energy densities for SkX, SC, and SP states, and the analytical energy density for the FM phase.

\subsection{Rashba SOC}
\begin{figure}[!htb]
\includegraphics[width=0.98\columnwidth]{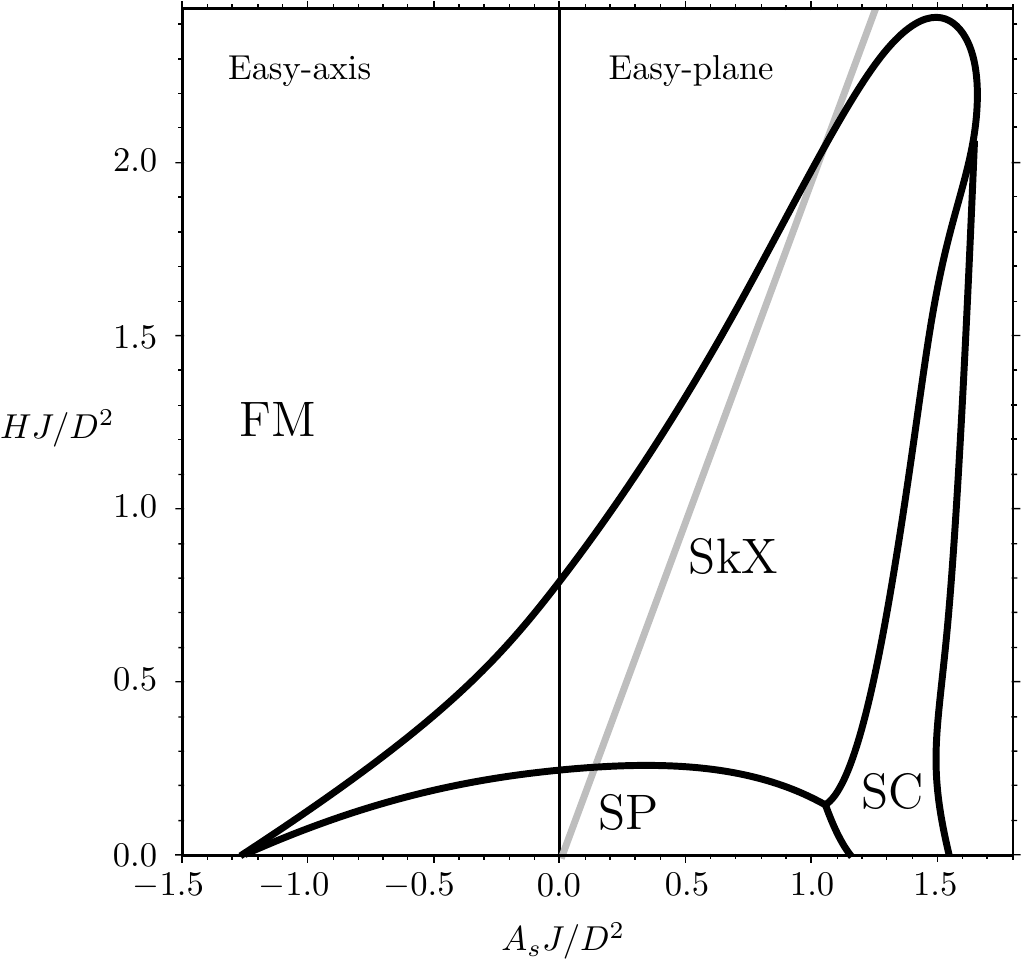}
\caption{Zero temperature phase diagram for the pure Rashba or pure Dresselhaus type symmetry is obtained by numerically solving the LLG equation. The same phase diagram also applies to the case of SO(2) symmetric DM tensor, $\hat D =  -D \openone$.  The grey line separates the aligned and the tilted regions of the FM phase. This phase is taken over by SkX, SP, and SC phases in the regions defined by the bold lines.}
\label{fig:Rashba}
\end{figure}

\begin{figure}[htbp]
\centering
\includegraphics[width=0.46\columnwidth]{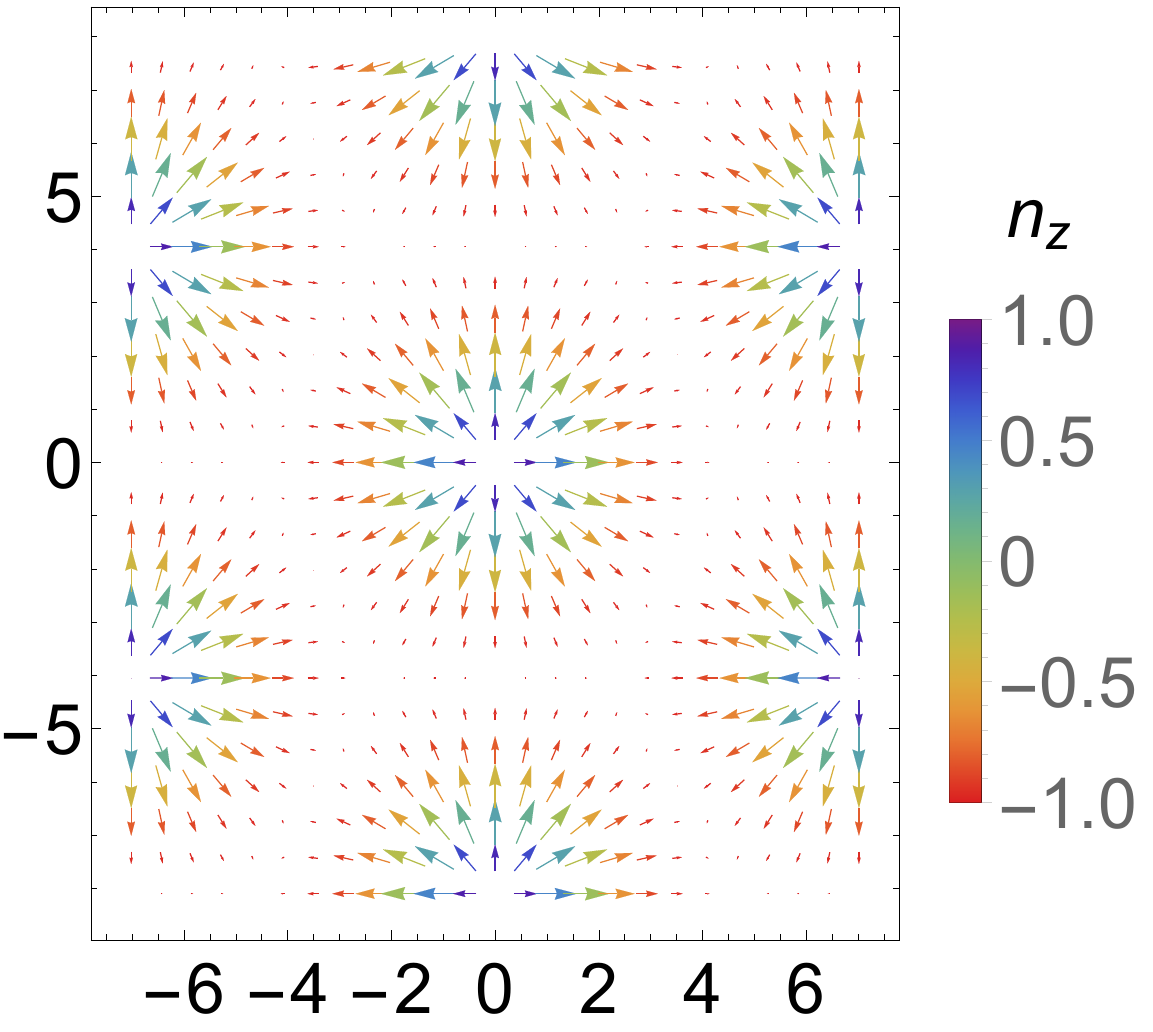}\hskip 0.1in
\includegraphics[width=0.48\columnwidth]{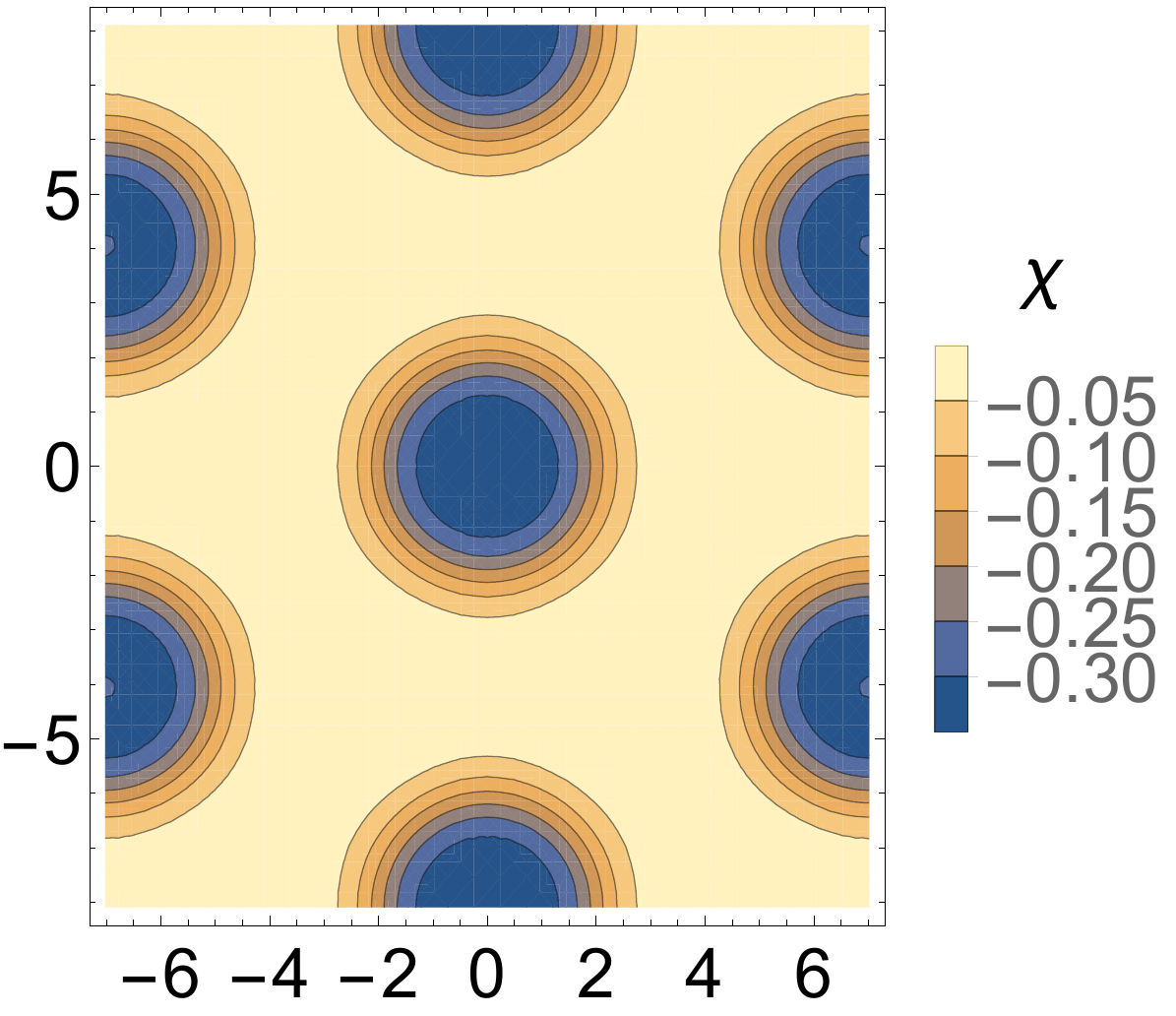}\\
\includegraphics[width=0.46\columnwidth]{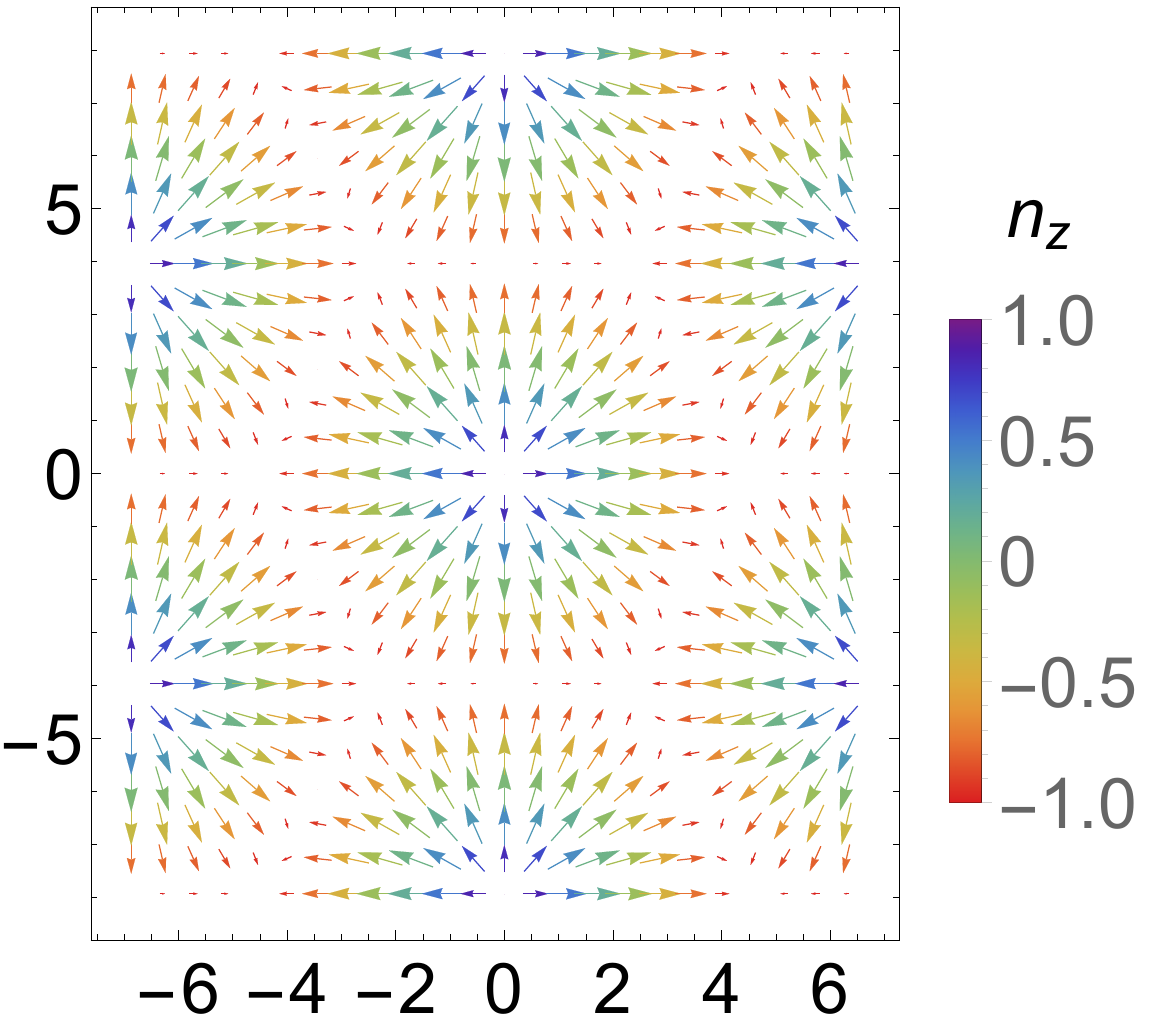}\hskip 0.1in
\includegraphics[width=0.48\columnwidth]{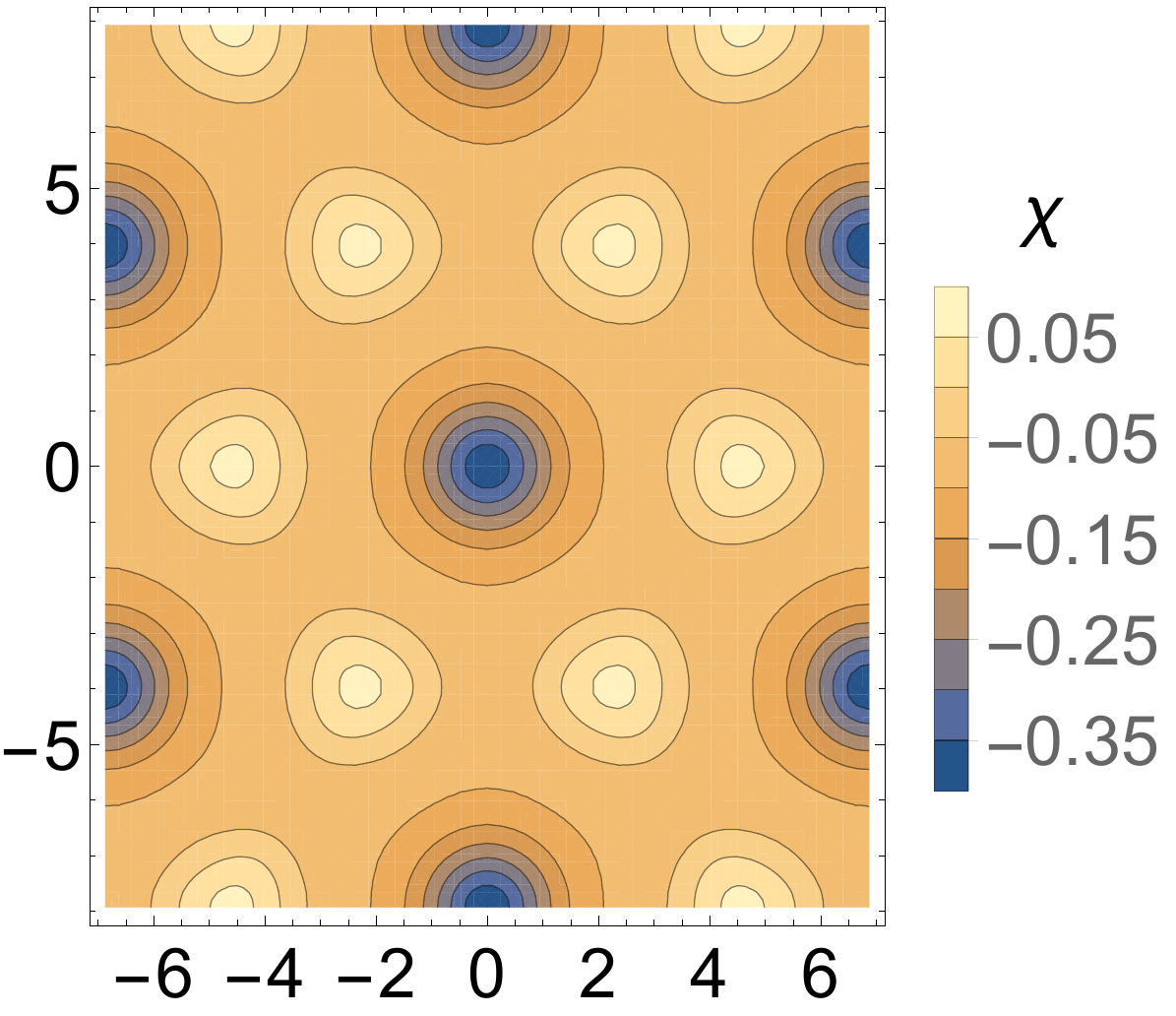}
\caption{(Color online) Normalized spin density $\boldsymbol n$ and topological charge density $\chi$ in SkX phase for the Rashba SOC. The size and direction of the arrows represent the in-plane component of $\boldsymbol n$ and the color represents $n_z$. $A_s J/D^2=0$ and $H J/D^2=0.7$  (top), $A_s J/D^2=0.8$ and $H J/D^2=0.7$ (bottom). As the strength of the easy-plane anisotropy is increased, localized skyrmions undergo a continuous charge splitting. When the relative size of the antivortices reach a critical value, square packing becomes energetically favorable and a first-order phase transition occurs into SC phase.}
\label{fig:RashbaSkX}
\end{figure}

\begin{figure}[htbp]
\centering
\includegraphics[width=0.47\columnwidth]{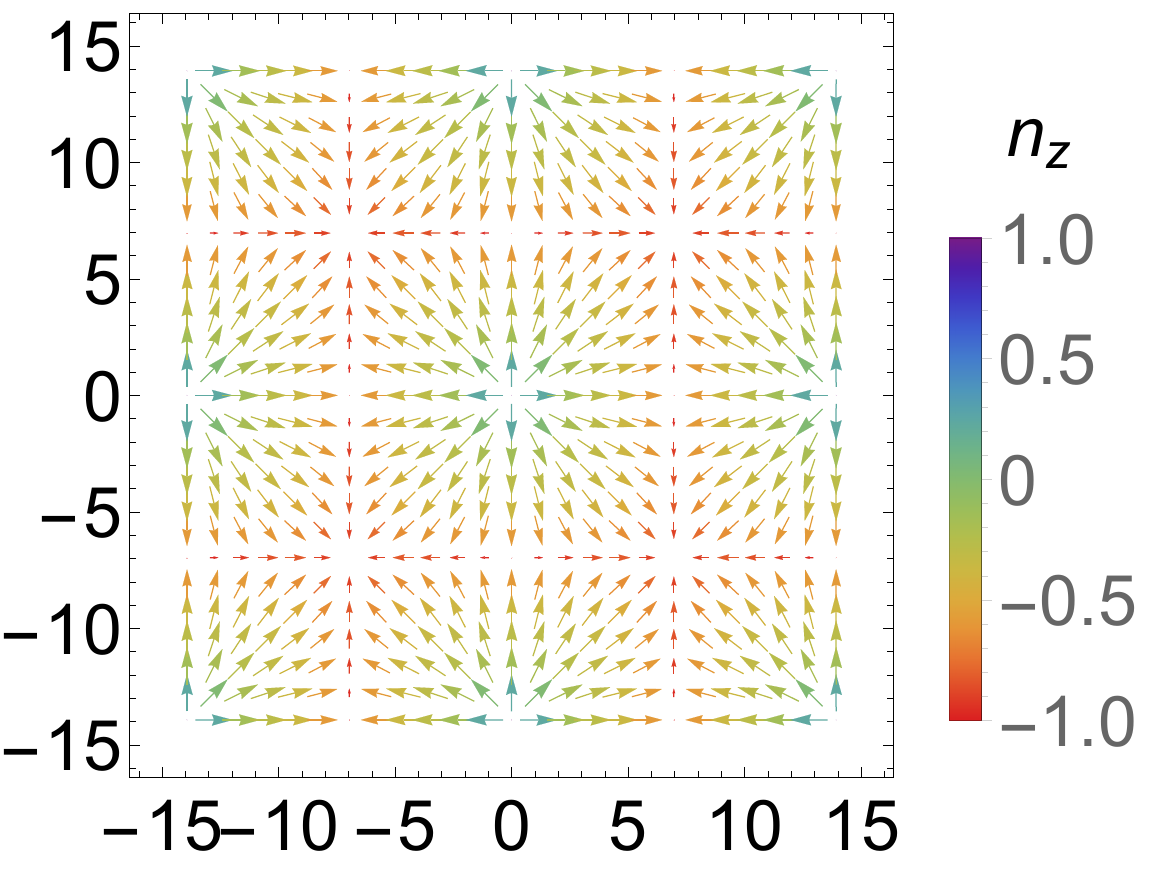}\hskip 0.1in
\includegraphics[width=0.45\columnwidth]{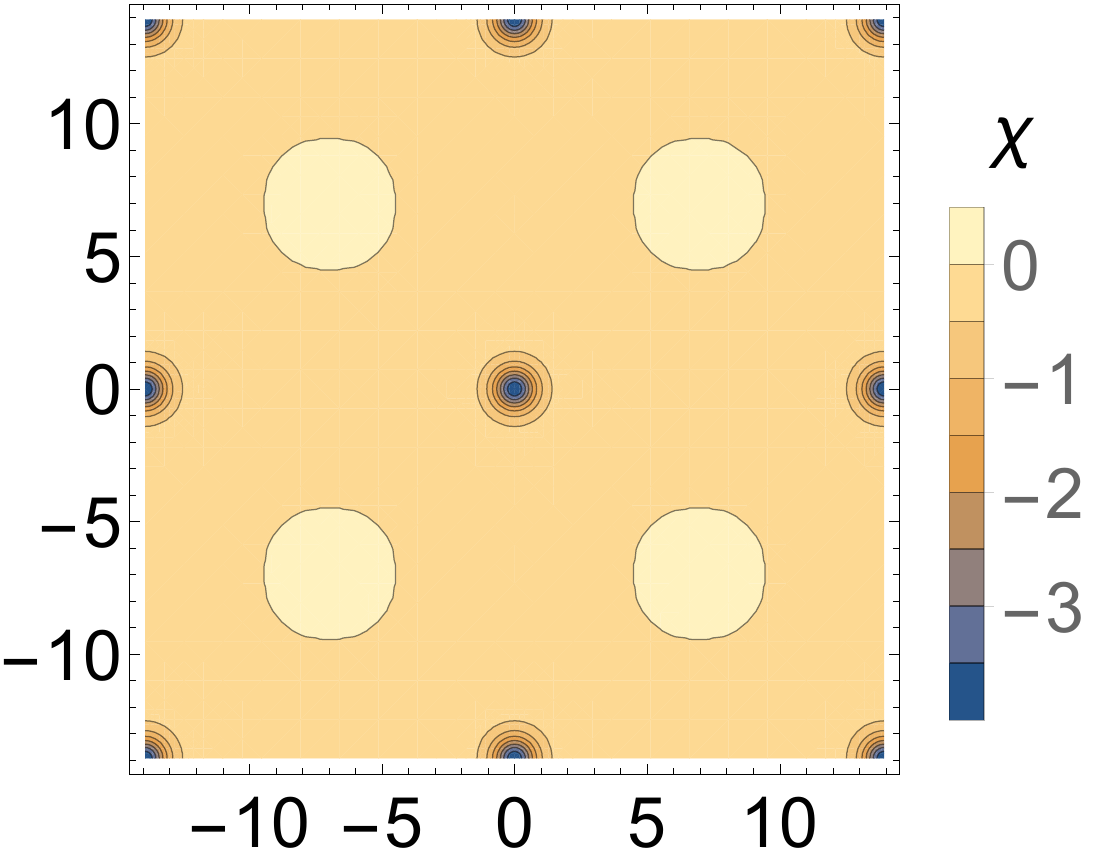}
\caption{(Color online) Normalized spin density and topological charge density at $A_s J/D^2=1.5$ and $HJ/D^2=1.4$ (SC phase) for Rashba SOC.}
\label{fig:RashbaSC}
\end{figure}
Here we study the case of the DM tensor given by $\hat D = -D \hat J_z$. The cases $\hat D = -D \hat J_z$ and $\hat D = -D \openone$ (which are equivalent to each other up to a global transformation, as discussed in the previous section) have been studied in Refs.~\onlinecite{Banerjee2014,Lin2015}. Our zero-temperature phase diagram shown in Fig.~\ref{fig:Rashba} mostly agrees with \cite{Banerjee2014}, except for the additional SC region which was missing in their analysis. SkX phase shows hedgehog-like skyrmions with well-localized topological charge $Q=1$ (Fig.~\ref{fig:RashbaSkX}). The same phase diagram also applies to the pure Dresselhaus case, $\hat D = -D \hat \lambda_1$, but skyrmions have $Q=-1$ due to the reflection involved in the equivalence transformation, as can be seen from Eq.~(\ref{eq:reflection}). As the easy-plane anisotropy is increased the topological charge of skyrmions gradually splits, forming the precursor to the vortex-antivortex pair lattice shown in Fig.~\ref{fig:RashbaSC}, while the total charge within a single unit cell remains $Q=1$.
As pointed out in \cite{Lin2015} the core radius of skyrmions in the SkX phase becomes larger and skyrmions start to overlap causing the formation of vortices/antivortices during this process. When the relative size of antivortices (with respect to vortices) reaches a critical value, the square lattice becomes the energetically more favorable packing for the vortex-antivortex lattice and a first-order phase transition occurs from SkX phase to SC phase.

\subsection{Rashba + Dresselhaus SOC}
\begin{figure}[!htb]
\includegraphics[width=0.98\columnwidth]{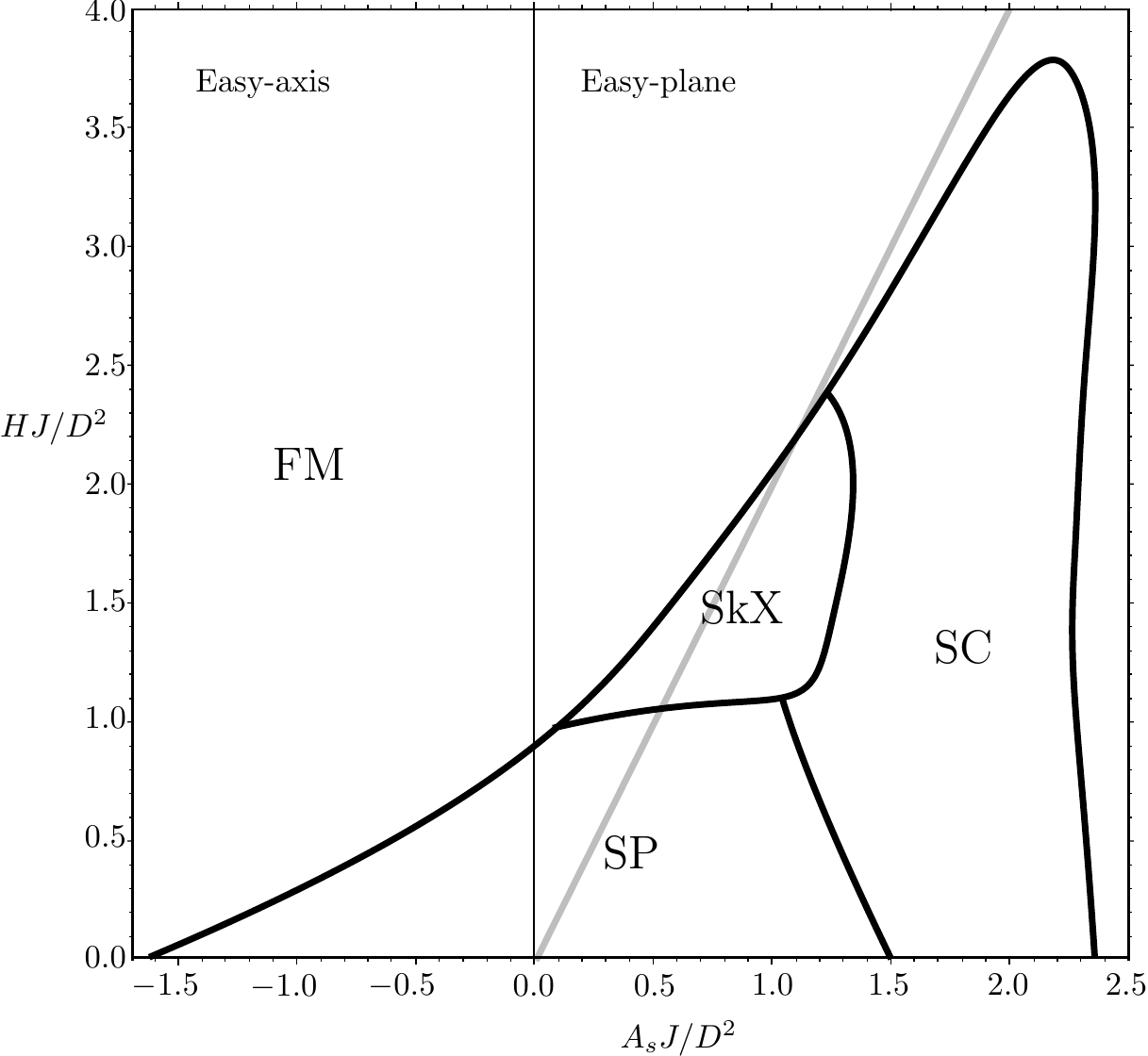}
\caption{Zero temperature phase diagram for the Rashba + Dresselhaus SOC with $C_{2v}$ symmetry ($\hat D = -D_R \hat J_z - D_D \hat \lambda_1$ with $D_R/D_D = 5$). SkX phase is only present in the easy-plane region ($A_s J/D^2>0$) of the phase diagram. The grey line separates the aligned and the tilted regions of the FM phase \cite{Clarke2007}, whereas SkX and SP phases are not affected by this line.}
\label{fig:C2v}
\end{figure}

\begin{figure}[htbp]
\centering
\includegraphics[width=0.48\columnwidth]{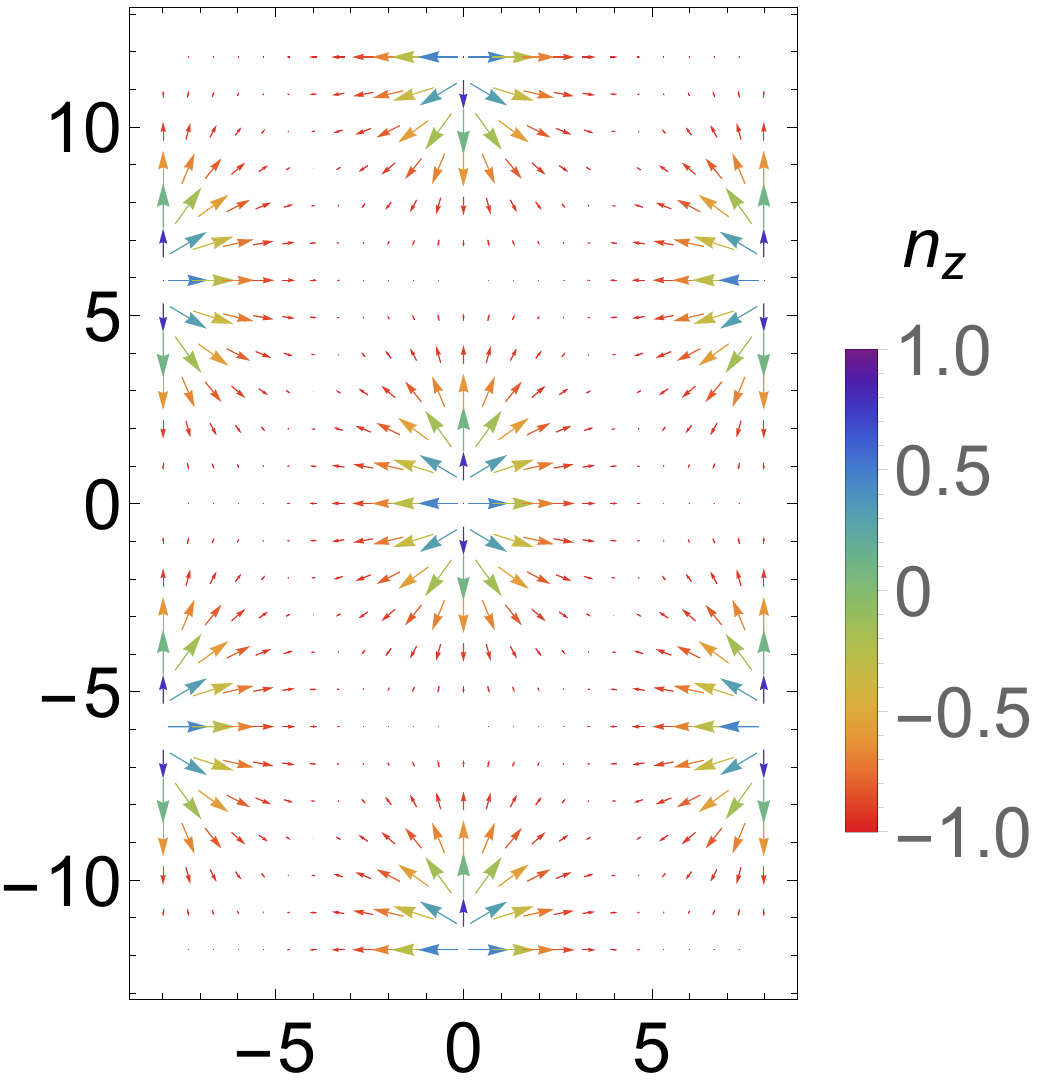}\hskip 0.1in
\includegraphics[width=0.48\columnwidth]{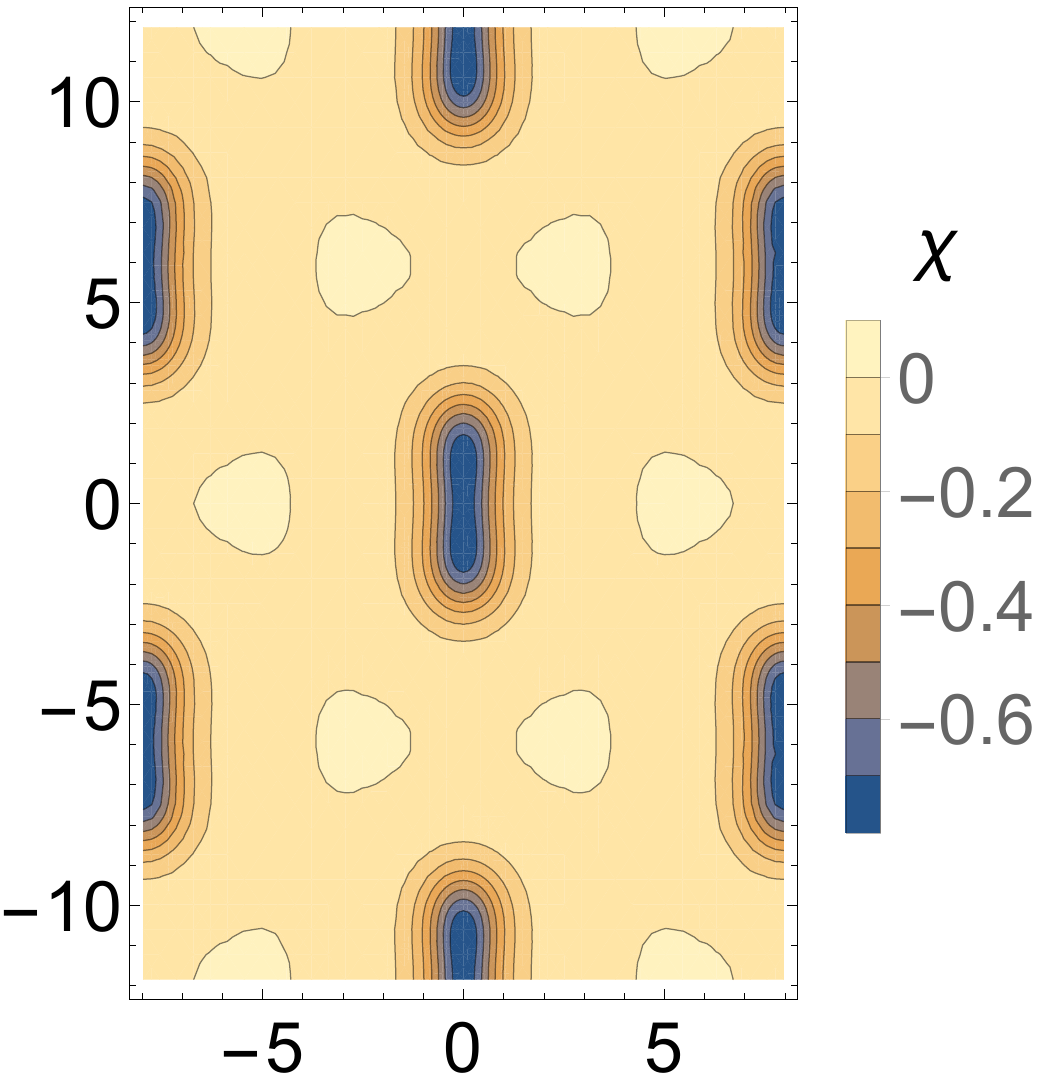}
\caption{(Color online) Normalized spin density and topological charge density at $A_s J/D^2=0.4, H J/D^2=1.2$ (SkX phase) for the Rashba + Dresselhaus SOC with $C_{2v}$ symmetry. This plot shows the initial stage of charge splitting of skyrmions elongated along the mirror planes ($xz$ and $yz$ planes).}
\label{fig:C2vSkX}
\end{figure}

\begin{figure}[!htb]
\centering
\includegraphics[width=0.48\columnwidth]{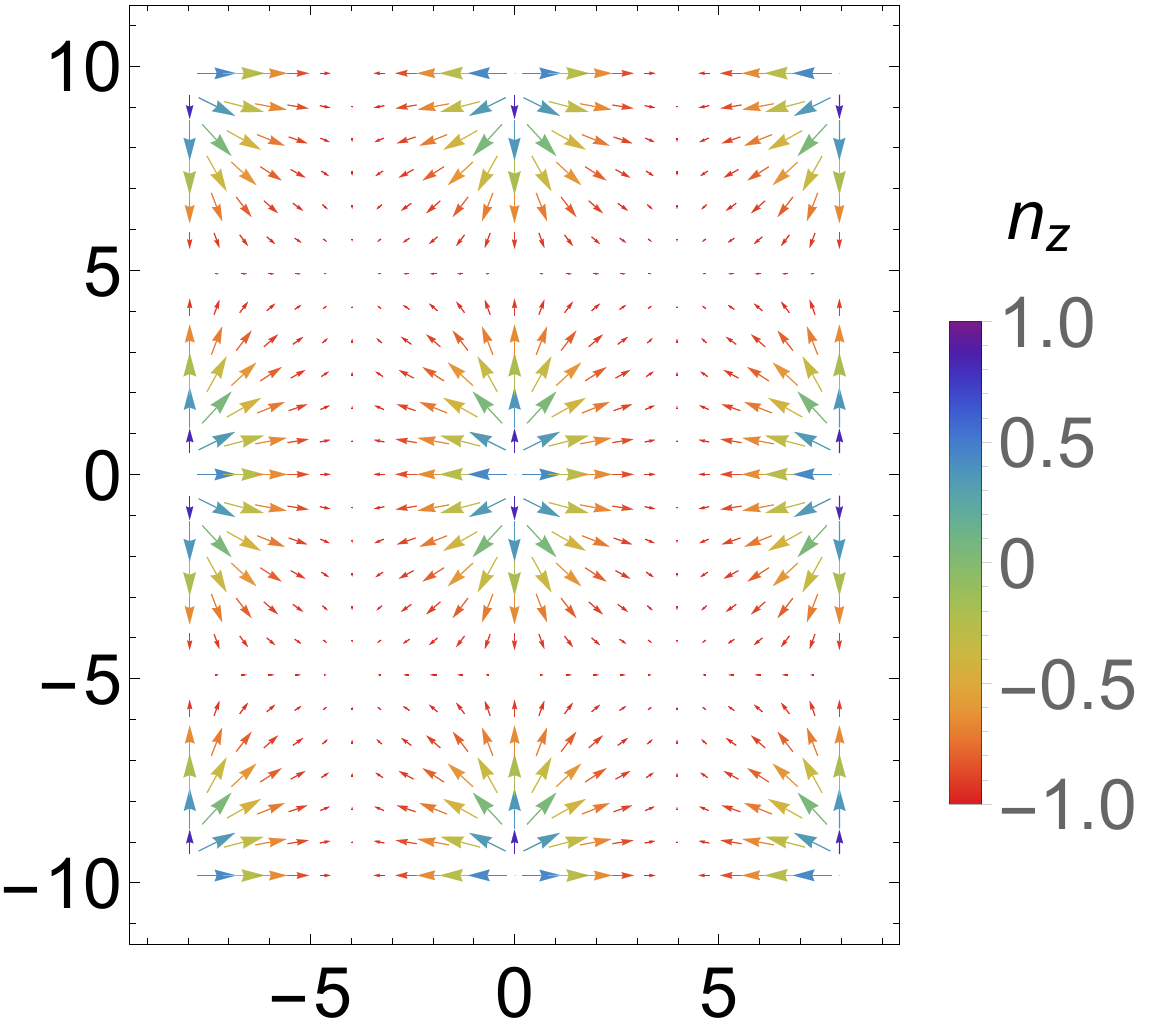}\hskip 0.1in
\includegraphics[width=0.48\columnwidth]{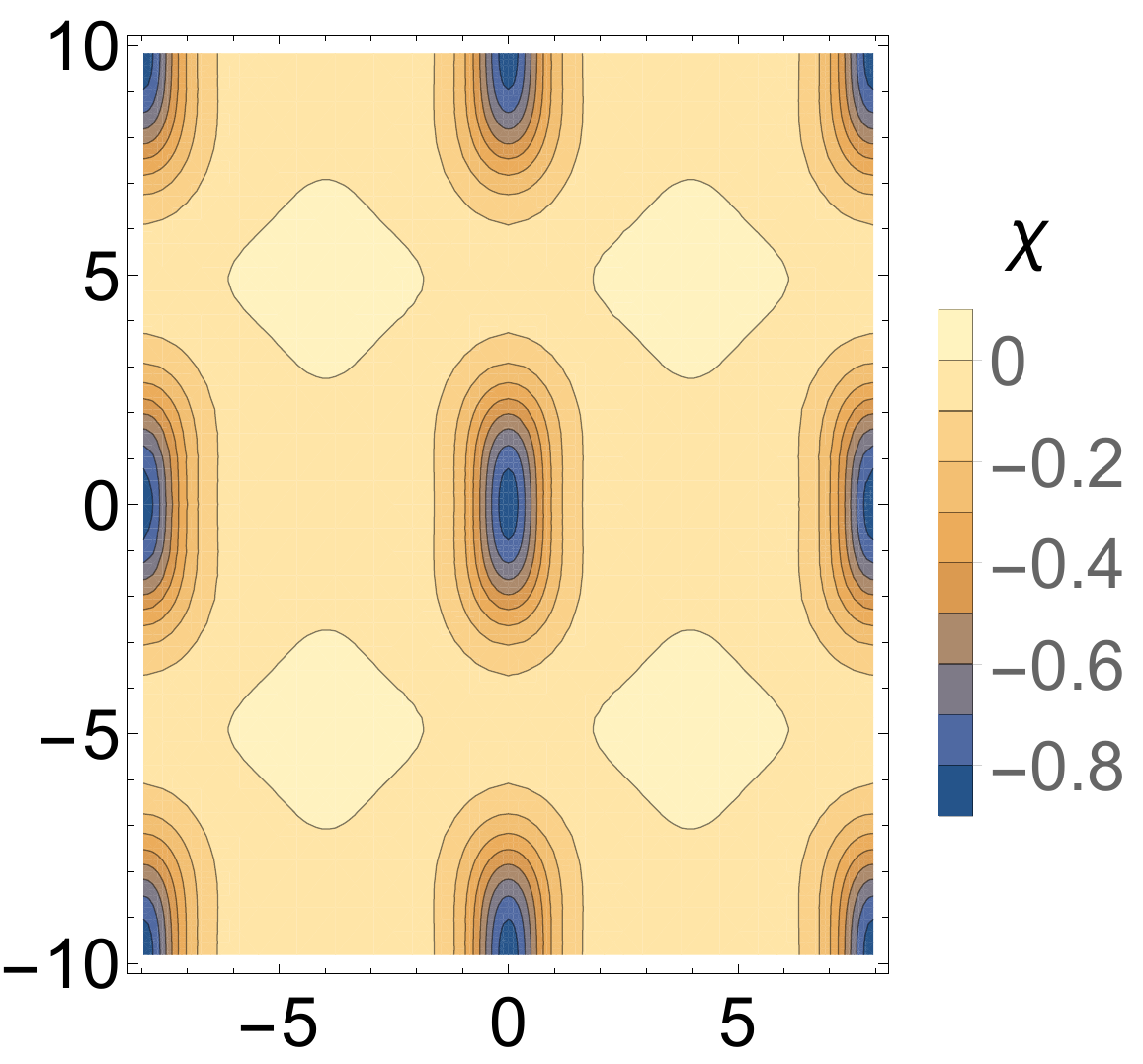}
\caption{(Color online) Normalized spin density and topological charge density at $A_s J/D^2=0.6, H J/D^2=1.3$ (SC phase) for the Rashba + Dresselhaus SOC with $C_{2v}$ symmetry. The vortices and antivortices are noticeably elongated.}
\label{fig:C2vSC}
\end{figure}

The mixture of Rashba and Dresselhaus SOC \cite{Ganichev2004,Ganichev2014,Oh2014} with $C_{2v}$ symmetry has been studied in \cite{Oh2014} in the absence of anisotropies, with the conclusion that SkX phase cannot exist when both SOC terms are present. Our analysis here shows that SkX phase can be stabilized by the uniaxial anisotropy. Figure~\ref{fig:C2v} shows the phase diagram for $\hat D = -D_R \hat J_z - D_D \hat \lambda_1$ with $D_R/D_D = 5$ ($D_R, D_D > 0$). The DM tensor of mostly of Rashba type with an additional small symmetry-breaking Dresselhaus type term results in skyrmions with the topological charge $Q=1$. In the opposite situation of $D_D/D_R = 5$, skyrmions converge to topologically-different Dresselhaus type skyrmions with $Q=-1$ charge. We observe that SkX region shrinks while SP and SC regions expand.
The $C_{2v}$ symmetry allows deformations along the axes of reflection ($x$- and $y$-axes in this case), but unlike the SP and SC configurations, the sixfold symmetry of SkX is not compatible with such deformations.
Elongation of skyrmions along the axes of symmetry is evident in Figs.~\ref{fig:C2vSkX} and \ref{fig:C2vSC}.

\subsection{Rashba SOC with tilting}
\begin{figure}[!htb]
\includegraphics[width=0.98\columnwidth]{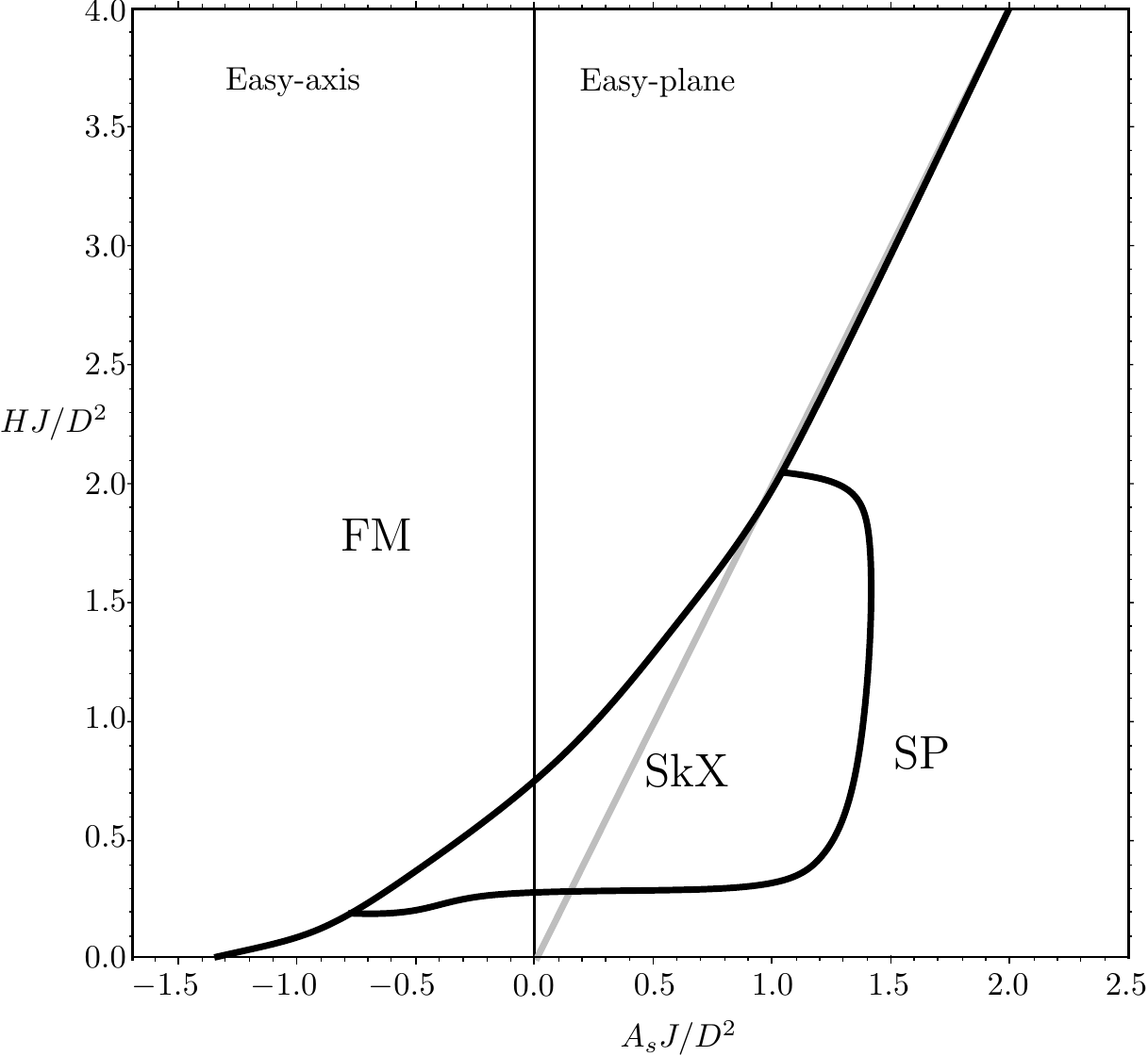}
\caption{Zero temperature phase diagram for the SOC with reflection symmetry ($\hat D = -D(0,\sin\theta_T,\cos\theta_T)^T \times$ with $\theta_T = \tan^{-1}(0.1) \approx 5.7^\circ$) which shows a SP phase with enhanced stability. The grey line separates the aligned and the tilted regions of the FM phase.}
\label{fig:Tilted}
\end{figure}

\begin{figure}[!htb]
\centering
%\raisebox{.015\height}{\includegraphics[width=0.48\columnwidth]{TiltedSkXMag}}\hskip 0.1in
\includegraphics[width=0.48\columnwidth]{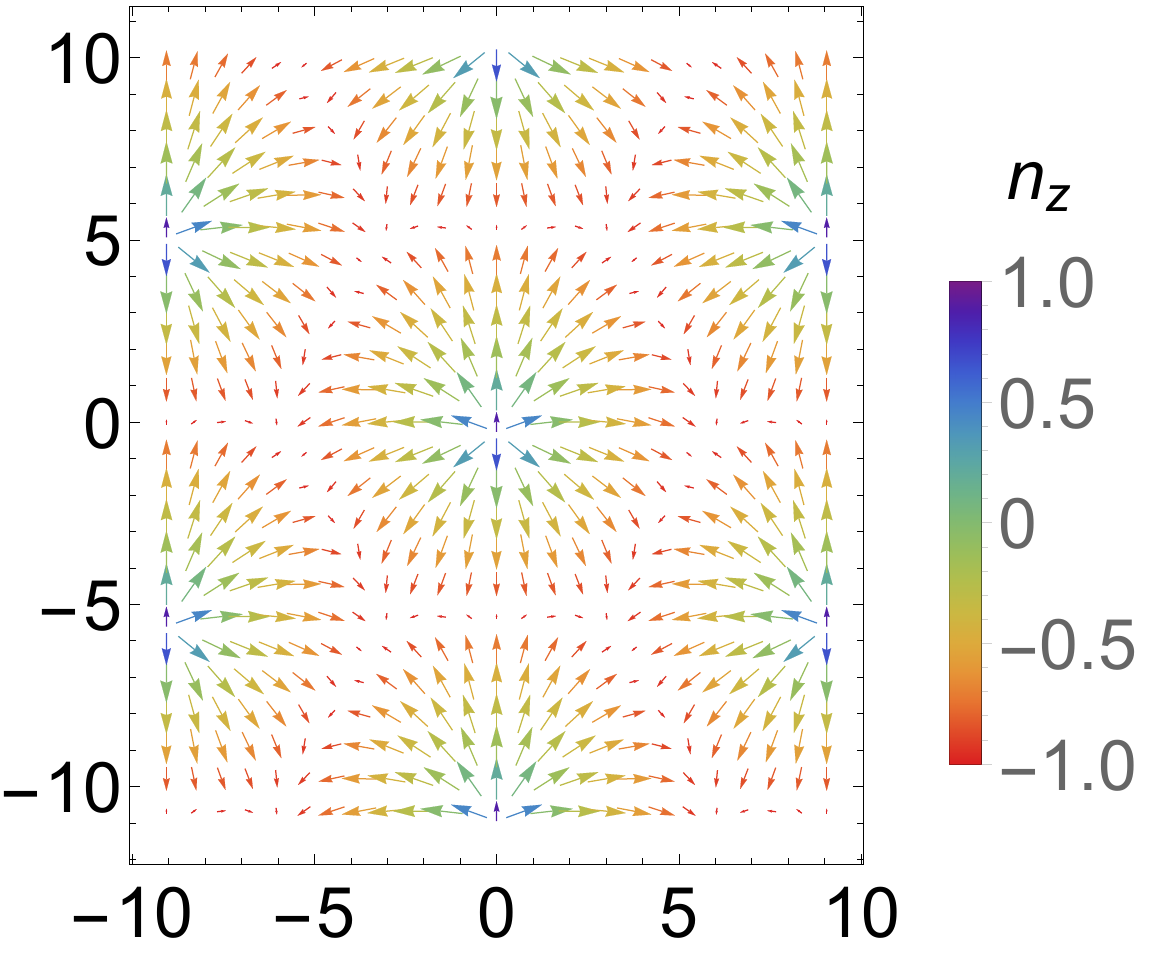}\hskip 0.1in
\includegraphics[width=0.46\columnwidth]{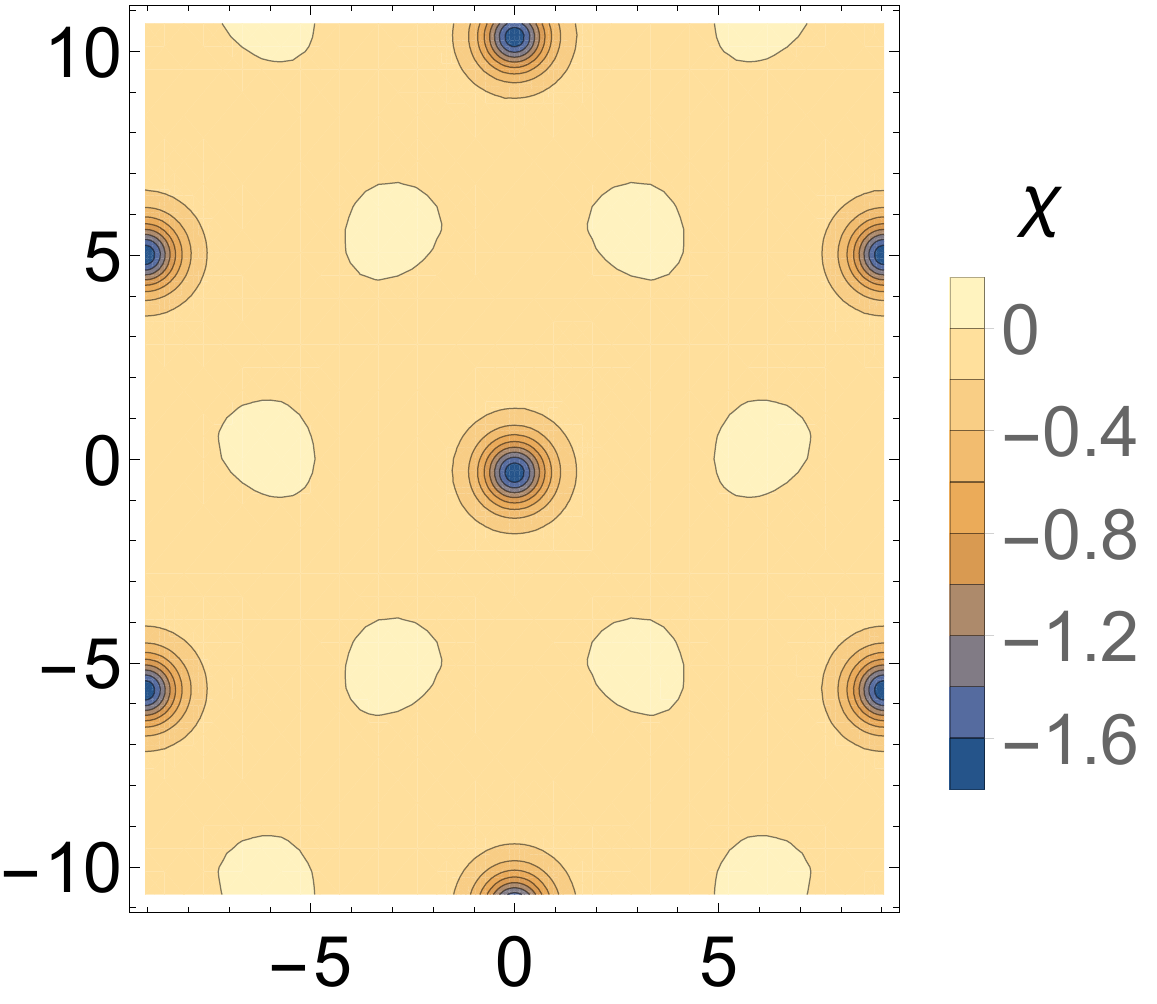}
\caption{(Color online) Normalized spin density and topological charge density at $A_s J/D^2=1.3, H J/D^2=1.5$ (SkX phase) for reflection symmetry. The core of skyrmions is shifted (from the unit cell center which is at the origin) along the axis of reflection.}
\label{fig:TiltedSkX}
\end{figure}

\begin{figure}[!htb]
\centering
\includegraphics[width=0.8\columnwidth]{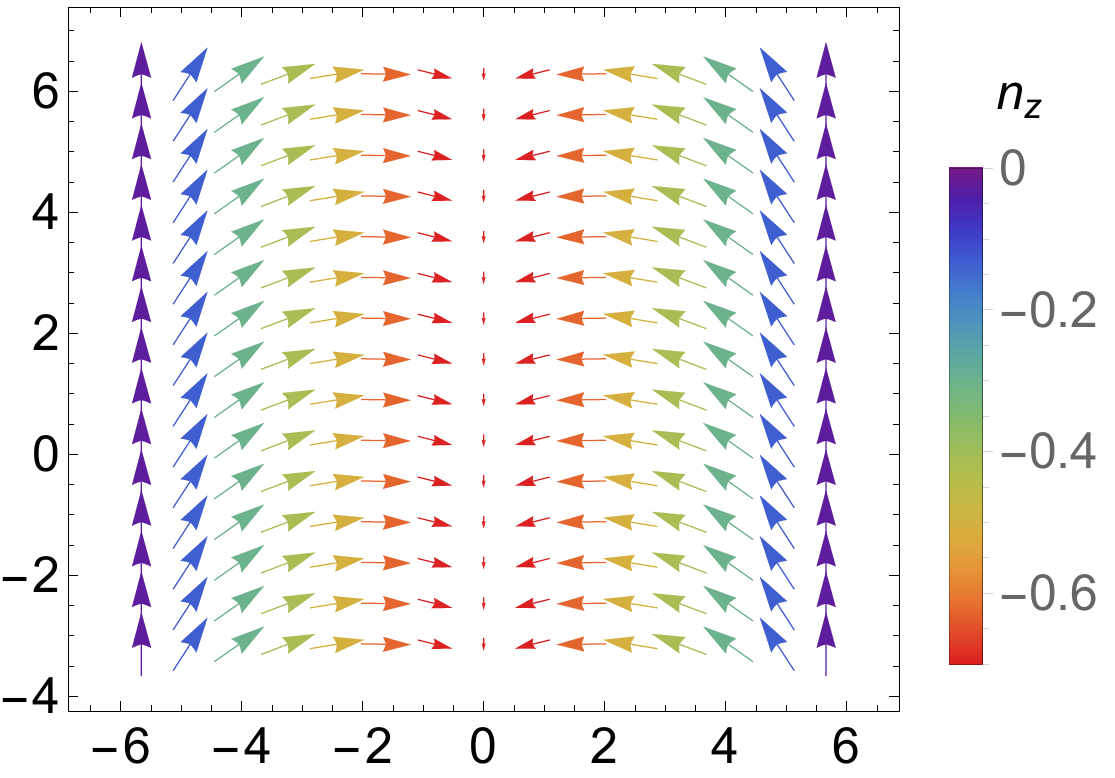}
\caption{(Color online) Normalized spin density at $A_s J/D^2=1.5, HJ/D^2=1$ (SP phase) for reflection symmetry, showing a mostly-in-plane spiral configuration. When the easy-plane anisotropy $A_s J/D^2$ is increased, spirals become almost completely in-plane.}
\label{fig:TiltedSP}
\end{figure}

Finally, we discuss the case in which the DM tensor is antisymmetric and corresponds to a vector that is making a small tilting angle with the $z$-axis: $\hat D = -D(0,\sin\theta_T,\cos\theta_T)^T \times$ and the tilting angle is chosen to be $\theta_T = \tan^{-1}(0.1) \approx 0.1$ or $5.7^\circ$. Such a system has reflection symmetry along a single mirror plane (which is the $yz$-plane for this choice of parameters). We find that the SP region expands greatly while the SkX region slightly shrinks and the SC region is completely replaced by SP, as can seen in Fig.~\ref{fig:Tilted}. It turns out that the SP phase also takes over the tilted FM region (in which the spin density is $n_z = -H/2A_s$) due to the fact that  such form of DM interaction favors in-plane spirals (Fig.~\ref{fig:TiltedSP}) and this allows exchange and DM interactions to lower the average energy density from $F_\text{FM}/A = A_s n_z^2 + H n_z$ (note that the in-plane component of the spin density does not contribute at all) where $F_\text{FM}$ is the free energy and $A$ is the area of the system.
The core of resulting skyrmions is shifted along the $y$-axis as can be seen in Fig.~\ref{fig:TiltedSkX}. There is also a slight elongation along the $y$-axis (around 2\% for the parameters used in the figures).

\section{Dynamics of skyrmions induced by spin currents}
Here we study the dynamics of skyrmions in response to spin currents. Spin currents naturally arise in a conducting ferromagnet in the presence of charge currents. Spin currents can also arise in a ferromagnetic insulator in a form of  magnon current as a response to a temperature gradient as we show in Appendix \ref{sec:LLG} by deriving the LLG equation from the stochastic LLG equation. In both cases, we can apply the following LLG equation:
\begin{align}
s (1+ \alpha \boldsymbol n_s \times) \dot {\boldsymbol n_s} = \boldsymbol n_s \times \boldsymbol H_\text{eff} - (1 +\beta \boldsymbol n_s \times) (\boldsymbol j^s \cdot \boldsymbol {\mathcal D} ) \boldsymbol n_s ,
\label{eq:LLG}
\end{align}
where $s$ is the local spin density, $\beta$ is the dissipative correction to the magnonic torque, $\boldsymbol j^s$ is the effective spin current induced either by charge carriers or by magnon currents, $\boldsymbol {\mathcal D}$ = $(\mathcal D_x, \mathcal D_y, \mathcal D_z)$ is the chiral derivative \cite{Kim2013a,Tserkovnyak2014}, and $\boldsymbol n_s$ represents the spin density whose dynamics is determined by the external magnetic field, magnetic anisotropies, and spin currents. 
The form of the chiral derivative is determined by symmetries of the system. In the most simple case it can be determined by DM interactions, i.e., $\mathcal D_\mu = \partial_\mu + (\hat D \boldsymbol e_\mu/J) \times$ (see Appendix \ref{sec:LLG} for a detailed discussion for the case of magnon currents and \cite{Matos-Abiague2009} for charge currents). However, in the most general settings the tensor involved in the chiral derivative can be renormalized, e.g., due to various scattering processes.

For describing the motion of skyrmions, we use Thiele's approach \cite{Thiele1973,Tretiakov2008,Schulz2012a, Nagaosa2013,Tomasello2014,Iwasaki2013,Schutte2014} in which the motion of magnetic textures is constrained to a subspace described by the generalized (collective) coordinates $q_i$ in the form $\boldsymbol n_s = \boldsymbol n_s(\boldsymbol r-\boldsymbol q(t))$. Under this assumption, the magnetic structure drifts as a whole while maintaining its internal structure: $\dot {\boldsymbol n}_s$ becomes $-\sum_\mu \dot q_\mu \partial_{q_\mu} \boldsymbol n_s$ and the equations of motion for $q_\mu$ can be found by applying the operator $\int_\text{cell} d^2 \boldsymbol r (\partial_{q_\mu} \boldsymbol n_s) \cdot \boldsymbol n_s \times$ to the LLG equation~(\ref{eq:LLG}). We obtain the following equation for the generalized coordinate:
\begin{align}
  s( \hat G + \alpha \hat \eta) \boldsymbol v  -(\hat G_1 + \beta \hat \eta_1) \boldsymbol j^s = 0 ,
\end{align}
where $\boldsymbol v = \dot {\boldsymbol q}$ denotes the speed of the skyrmion and
\begin{align}
[\hat G]_{\mu \nu} =& \frac{1}{4 \pi}\int d^2 \boldsymbol r  (\partial_\mu \boldsymbol n_s \times \partial_\nu \boldsymbol n_s) \cdot \boldsymbol n_s = \int d^2 \boldsymbol r \chi(\boldsymbol r) \epsilon_{\mu \nu z}, \nonumber \\
[\hat G_1]_{\mu \nu} =& \frac{1}{4 \pi}\int d^2 \boldsymbol r  (\partial_\mu \boldsymbol n_s \times \mathcal D_\nu \boldsymbol n_s) \cdot \boldsymbol n_s, \nonumber \\
[\hat \eta]_{\mu \nu} =& \frac{1}{4 \pi}\int d^2 \boldsymbol r  (\partial_\mu \boldsymbol n_s \cdot \partial_\nu \boldsymbol n_s), \nonumber \\
[\hat \eta_1]_{\mu \nu} =& \frac{1}{4 \pi}\int d^2 \boldsymbol r  (\partial_\mu \boldsymbol n_s \cdot \mathcal D_\nu \boldsymbol n_s),
\label{eq:thiele-tensors}
\end{align}
with $\mu,\nu \in \{x,y\}$ and the integrations are over a single unit cell. 
The antisymmetric gyrotensor $\hat G$ can be written as $-Q \boldsymbol z \times$, and we have $\hat G = \hat G_1$. The damping dyadic tensors $\hat{\eta}$ and $\hat{\eta}_1$ account for the effects of dissipation. 

\begin{figure}[htbp]
\centering
\includegraphics[width=0.7\columnwidth]{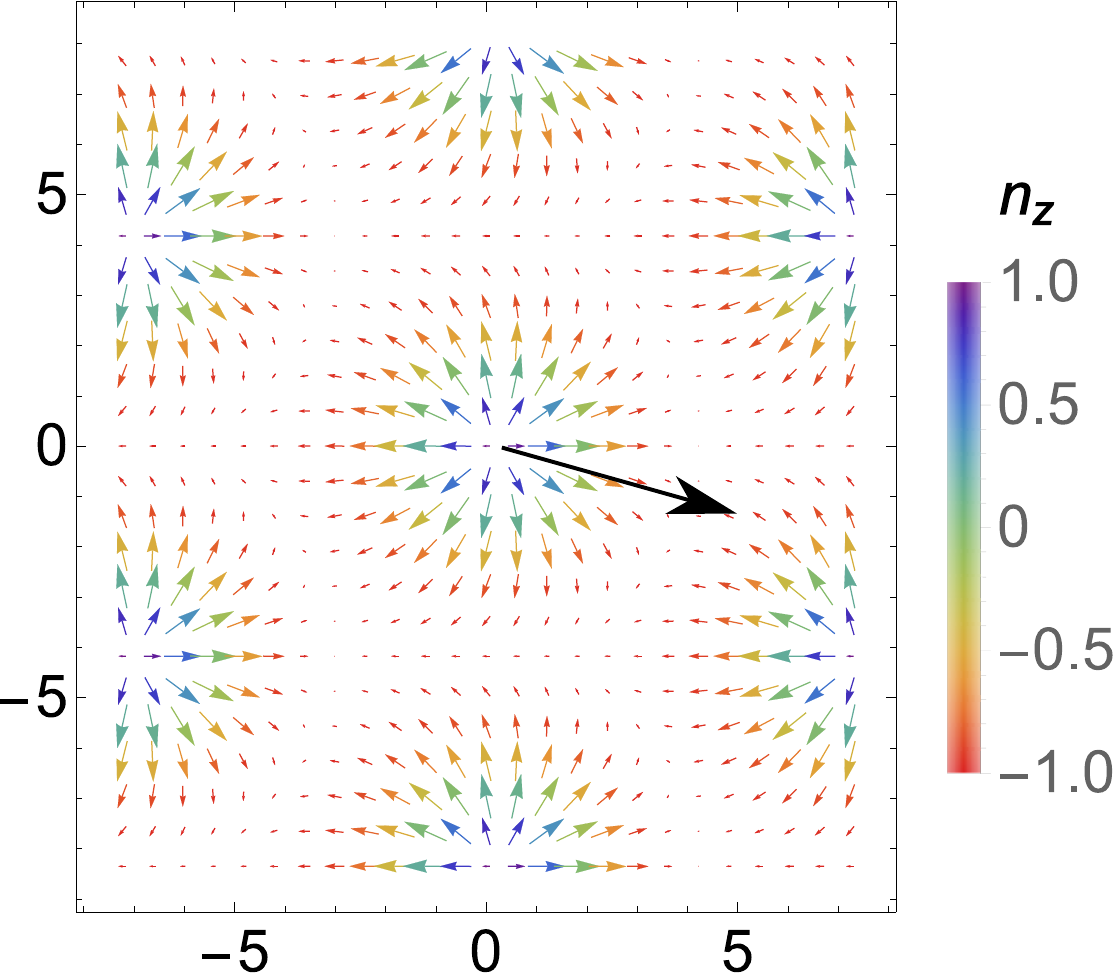}
\caption{(Color online)  Motion of $Q=1$ skyrmions due to spin current along the $x$-axis, simulated by the LLG equation with the torque term, Eq.~(\ref{eq:LLG}). Skyrmions move along the spin current (e.g. toward the hotter region for the case of magnon-mediated torques) ($+x$ direction) with an additional side motion ($-y$ direction). Skyrmions get deformed along the direction of motion.
}
\label{fig:MagCurrent}
\end{figure}

For the rotationally symmetric case a skyrmion of radius $R$ will result in $\hat \eta=\eta \openone$ and $\hat \eta_1=\eta_1 \openone$, with
\begin{align}
\eta &= \pi \int_0^R dr \left( \frac{\sin^2 n_\theta}{r} + r (\partial_r n_\theta)^2 \right), \nonumber \\
\eta_1 &= \eta + \pi \int_0^R dr (\sin n_\theta \cos n_\theta + r \partial_r n_\theta),
\end{align}
where $n_\theta$ and $n_\phi$ are the spherical coordinates of $\boldsymbol n$, leading to the equation of motion
\begin{align}
 \left(\eta  \alpha s \boldsymbol v -  \eta_1 \beta \boldsymbol j^s\right)  -Q \boldsymbol z \times (s \boldsymbol v -\boldsymbol j^s) = 0.\label{eq:coll}
\end{align}
For the motion of skyrmions in response to a time-independent spin current along $x$-direction ($\boldsymbol j^s = j^s \boldsymbol e_x$), this equation yields
\begin{align}
v_x = j^s\frac{ Q^2 + \alpha \beta \eta \eta_1 }{s( Q^2 + \alpha^2 \eta^2)}, \quad v_y = j^s  Q \frac{\beta \eta_1-\alpha\eta}{s( Q^2 + \alpha^2 \eta^2)},\label{eq:vel}
\end{align}
that is, skyrmions will move along the spin current with an additional side motion, resulting in a Hall-like motion with Hall angle $\theta_H = \tan^{-1}(v_y / v_x)$ (see Fig.~\ref{fig:MagCurrent}).

In cases lacking the rotational symmetry, the dissipation tensor can be diagonalized as $\hat \eta' = \hat R_z \hat \eta \hat R_z^T = \text{diag}(\eta_{xx}',\eta_{yy}')$ by a proper rotation $\hat R_z$ around the $z$-axis that aligns the basis vectors with the preferred directions due to the broken symmetry. $\hat R_z$ also diagonalizes $\eta_1'$. Such a transformation does not affect $\hat G$ since $[\hat R_z, \hat J_z]=0$, thus the equation of motion in the new coordinate system becomes
\begin{align}
 \hat \eta' \alpha s \boldsymbol v' - \hat \eta_1' \beta {\boldsymbol j^s}' -Q \boldsymbol z \times (s \boldsymbol v'-{\boldsymbol j^s}') = 0,
\end{align}
or
\begin{align}
Q \boldsymbol z \times (s \boldsymbol v'-{\boldsymbol j^s}') - 
\begin{pmatrix}
\eta'_{xx} & 0 \\
0 & \eta'_{yy}
\end{pmatrix} \alpha s \boldsymbol v' \nonumber \\
+\begin{pmatrix}
{\eta_1'}_{xx} & 0 \\
0 & {\eta_1'}_{yy}
\end{pmatrix} \beta {\boldsymbol j^s}' = 0,
\end{align}
where ${\boldsymbol j^s}' = \hat R_z \boldsymbol j^s$ and $\boldsymbol v' = \hat R_z \boldsymbol v$.

In SkX phase, inter-skyrmion interactions force the inter-skyrmion distance to a particular value. In the case when chiral derivative is given by $\mathcal D_\mu = \partial_\mu + [\hat D \boldsymbol e_\mu/J] \times$  this leads to $\hat \eta_1' =0$ and $\theta_H = -Q \alpha \eta$ (see Appendix \ref{sec:scaling}). However, this exact cancellation does not happen for isolated skyrmions in general and when the chiral derivative entering the LLG equation (\ref{eq:LLG}) is renormalized. Nevertheless, we observe that the renormalization of the dissipative tensor in Eq.~(\ref{eq:thiele-tensors}) has to be taken into account, especially for skyrmion lattices. Such renormalization was not considered in the previous studies \cite{Fert2013,Iwasaki2013,Lin2014,Lin2015}.

Isolated skyrmions in a chiral magnet can be realized by increasing the magnetic field in SkX phase or by injection of spin currents. Such isolated skyrmions will exist as topologically protected defects \cite{Ler2010,Butenko2010,Leonov2015}, whose Hall motion is affected by the $\beta$-term as well (see e.g. Eq.~(\ref{eq:coll})). On the other hand, for isolated skyrmions the inter-skyrmion distances become much larger compared to the skyrmion core size, for which we estimate $\hat \eta_1 \approx \hat \eta$. In the particular case of $\alpha = \beta$   (this case is realized for magnon-mediated torques for $d=2$, see Appendix \ref{sec:LLG}), isolated skyrmions will move along the spin current without a side motion (see Eq.~(\ref{eq:vel})), similar to antiferromagnetic skyrmions \cite{Barker2015}. This can have implications for realizations of magnetic memories relying on skyrmions for information encoding. 

While the form of the SOC completely determines the helicity $\gamma$ and topological charge $Q$ of the skyrmions in SkX phase, in principle it is possible to create metastable skyrmions with different helicity $\gamma'$  and charge $Q'$. The dynamics of such skyrmions will be different in general: $\boldsymbol n$ in Eq.~(\ref{eq:thiele-tensors}) will be replaced by $\hat R' \boldsymbol n$, where $\hat R' = \hat R_z(\gamma'-\gamma) (-\lambda_3)^{(1- Q Q')/2}$, which in turn means $\hat G \to \text{det}(\hat R') \hat G$, $\hat G_1 \to \text{det}(\hat R') \hat G_1$, $\eta \to \eta$ and $\eta_1 \to \hat R'^T(\eta_1-\eta) + \eta$.

Note that we consistently keep only the first order terms in SOC strength and assume a smooth magnetic texture. In sharp textures, there could be additional damping terms of the order of $(D/J)^2$ whose overall effect is to renormalize the elements of $\hat\eta$ \cite{Kim2015a,Zang2011}. Similar corrections can occur when the magnon wavelength is comparable to the texture size \cite{Schroeter2015}.

\section{Conclusions}
In this work, we have studied SkX and SC crystals at temperatures that are much lower than the Curie temperature.   Similar studies of magnetic skyrmions so far have been limited to systems with high symmtery. A previous study concluded that SkX phase does not exist in a system with both Rashba and Dresselhaus SOC, but uniaxial anisotropy was not present in their model \cite{Oh2014}. As we have shown, SkX phase is present even in systems with reduced symmetries; however, the skyrmions become asymmetric. In fact, we have established a clear connection between the symmetries of skyrmions and the corresponding DM interactions (see, e.g., Fig.~\ref{fig:skyrmions}). In addition, we have found that reduced symmetries result in enhanced stability of vortex-antivortex lattices and spirals, even in the absence of an external magnetic field. This behavior  has also been reported in \cite{Han2010} for MnSi ($\hat D = \openone$). In our Monte-Carlo simulations, we have observed anisotropy-driven, first order phase transitions between FM, SkX, SC, and SP phases.

We have also studied the dynamics of skyrmions and SkX (SC) lattices induced by spin currents where the spin current is induced by charge carriers or by magnons.  We have found striking differences between the motion of lattices and isolated skyrmions. This difference arises due to a renormalization of the dissipative dynamics of SkX and SC lattices and can be expressed via the chiral derivative. On the other hand, this renormalization is not important for isolated skyrmions. As a consequence,  our theory indicates that under certain conditions, isolated skyrmions can move along the current without a side motion, similar to antiferromagnetic skyrmions.  This can have implications for realizations of magnetic memories relying on skyrmions for information encoding.

Our results applies to monolayers as well as quasi-2D layers thinner than the pitch of out-of-plane spirals with the length scale $J/D$ \cite{Banerjee2013,Rowland2015,Yu2010,Wilson2014,Buhrandt2013}.

To conclude, our results relate different SkX and SC phases to the symmetries of materials used for realizations of skyrmions. This will give clear directions for experimental realizations of SkX and SC phases, and will allow engineering of skyrmions with unusual properties.

\begin{acknowledgments}
We thank Mohit Randeria for bringing to our attention Ref. \cite{Rowland2015}.
This work was supported in part by the DOE Early Career Award DE-SC0014189, and by the NSF under Grants Nos. Phy-1415600, PHY11-25915, and DMR-1420645.
O.A.T. acknowledges support by the Grants-in-Aid for Scientific Research (Nos. 25800184, 25247056, and 15H01009) from the Ministry of Education, Culture, Sports, Science and Technology (MEXT) of Japan and SpinNet.
K.B. acknowledges support from NSF Grant No. DMR-1308751.
\end{acknowledgments}

\appendix

\section{Dzyaloshinskii-Moriya interaction in itinerant ferromagnets}
\label{sec:chiral}
The discussion in this section is relevant to itinerant ferromagnets with DM interactions, e.g., thin magnetic films realized experimentally in Refs.~\cite{Dupe2014,Moreau-Luchaire2015}.
The effective free energy of a quasi-two-dimensional (2D) chiral magnet can be obtained starting from the 2D Rashba Hamiltonian with a general spin-orbit interaction term
\begin{align}
\hat H = \frac{\boldsymbol p^2}{2m} + (\hat \alpha \boldsymbol p) \cdot \boldsymbol \sigma.
\end{align}
Here the tensor $\hat{\alpha}$  describes SOC, $\boldsymbol{p}$ is the in-plane 2D electron momentum, $\boldsymbol \sigma = (\sigma_x, \sigma_y, \sigma_z)$. Note that SOC can be related to DM interactions by relation $\hat{\alpha}=\hbar/(2m J) \hat D$.
The Coulomb interactions resulting in the magnetic state are spin independent.Thus,  we can eliminate the spin-orbit interaction using a local SU(2) gauge transformation \cite{Kim2013a}
 $\hat U = \exp(-i \varphi \boldsymbol u \cdot \boldsymbol \sigma/2)$
to obtain
\begin{align}
\hat  U^\dagger \hat H \hat U &= \frac{p^2}{2m} + \mathcal O(\varphi^2),
\end{align}
where $\boldsymbol u = \hat D \boldsymbol r/|\hat D \boldsymbol r|$ is a unit vector along the axis of rotation, $\boldsymbol r$ is the in-plane position vector and the angle $\varphi = \hbar |\hat D \boldsymbol r| /J$ is proportional to the strength of the spin-orbit interaction. In the remainder of this Appendix, we will neglect higher order terms in $\varphi$ given that SOC is a relativistic effect.

The free energy density corresponding to the system in the rotated frame is given by $\mathcal F_0 = \sum_{\mu=x,y} J (\partial_\mu \bar {\boldsymbol n})^2/2$ where $J$ is the strength of the exchange interaction of the magnetic state with the order parameter $\bar{\boldsymbol{n}}$ originating from the Coulomb interactions.
The original and rotated frames are related to each other by an SO(3) rotation: $\boldsymbol n = \hat R \bar {\boldsymbol n}$, where $\hat R = \exp(\varphi \boldsymbol u \cdot \boldsymbol J)=\exp(\phi \boldsymbol u \times)$, $\boldsymbol J = (\hat J_x, \hat J_y, \hat J_z)$ and:
\begin{align}
\hat J_x = \begin{pmatrix}
0 & 0 & 0 \\
0 & 0 & -1 \\
0 & 1 & 0
\end{pmatrix}, 
\hat J_y = \begin{pmatrix}
0 & 0 & 1 \\
0 & 0 & 0 \\
-1 & 0 & 0
\end{pmatrix},
\hat J_z = \begin{pmatrix}
0 & -1 & 0 \\
1 & 0 & 0 \\
0 & 0 & 0
\end{pmatrix}
\label{eq:so3-generators}
\end{align}
are the generators of the fundamental representation of SO(3). ($\hat U$ and $\hat R$ represent the same rotation in spin- and real-space, respectively.) Since the gauge transformation is not a global one, we obtain a nontrivial covariant (often called chiral in this context) derivative $\mathcal D_\mu$,
\begin{align}
\mathcal D_\mu = \partial_\mu + \hat R \partial_\mu \hat R^T = \partial_\mu + (\hat D \boldsymbol e_\mu/J) \times,
\end{align}
which replaces the spatial derivative $\partial_\mu$ and captures the linear effects of SOC. The continuous free energy density in the original frame is then found to be
\begin{align}
\mathcal F_0 = \sum_{\mu=x,y} \frac{J}{2}(\mathcal D_\mu {\boldsymbol n})^2 \approx \sum_{\mu=x,y} \frac{J}{2} (\partial_\mu \boldsymbol n)^2 + (\hat D \boldsymbol e_\mu) \cdot (\boldsymbol n \times \partial_\mu \boldsymbol n).
\end{align}

\section{Tight-binding microscopic model}
\label{sec:microscopic}
In this Appendix, we derive the free energy based on a tight-binding microscopic model.
Such microscopic model can be relevant to realizations of magnetic systems at oxide interfaces \cite{Lee2013,Moetakef2012}
We describe our system by the tight-binding Anderson-Hasegawa Hamiltonian  $\hat H=\hat H_t + \hat H_\text{SO} + \hat H_\text{int}$ on a square lattice \cite{Banerjee2013}:
\begin{align}
\hat H_t &= -t \sum_{\langle i, j \rangle, \alpha} c^\dagger_{i\alpha} c_{j \alpha} + \text{H.c.}, \nonumber\\
\hat H_\text{SO} &=- i\lambda_\text{SO} \sum_{\langle i, j \rangle, \alpha \beta} c^\dagger_{i\alpha} c_{j \beta} \boldsymbol {\mathfrak D}_{ij} \cdot \boldsymbol \sigma_{\alpha \beta} + \text{H.c.}, \nonumber\\
\hat H_\text{int} &=  J\sum_{ij} \boldsymbol S_i \cdot \boldsymbol S_j - 2 J_H \sum_i \boldsymbol s_i \cdot \boldsymbol S_i,
\end{align}
where $\hat H_t$ is the nearest-neighbor hopping term, $\hat H_\text{SO}$ is the spin-orbit coupling and $\hat H_\text{int}$ is the interaction Hamiltonian, $J$ is superexchange energy, $J_H$ is Hund's coupling energy, $c_{i \alpha}^\dagger$ ($c_{i \alpha}$) creates (annihilates) an itinerant fermion with spin $\alpha$ at lattice site $i$, $\boldsymbol {\mathfrak D}_{ij} = \hat D \boldsymbol e_{ij}/D$, $\boldsymbol \sigma = (\sigma_x, \sigma_y, \sigma_z)$, $\boldsymbol e_{ij}$ is a unit vector from site $i$ to $j$, $\sum_{\langle i, j \rangle}$ denotes summation over nearest neighbors, $\boldsymbol s_i = \sum_{\alpha \beta} c_{i\alpha}^\dagger \boldsymbol \sigma_{\alpha \beta} c_{i\beta}/2$, and $\boldsymbol S_i$ denotes lattice-localized spins.

The SOC coupling can be gauged away by a rotation in the spin space, which rotates the first site by an angle $\phi_{ij}$ defined through $\tan\phi_{ij} = (|\hat D \boldsymbol e_{ij}|/D) (\lambda_\text{SO}/t)$ and the second site by $-\phi_{ij}$ around the axis $\boldsymbol u_{ij} = \hat D \boldsymbol e_{ij} / |\hat D \boldsymbol e_{ij}|$:
\begin{align}
\bar c_{i \alpha} = [\hat U^{ij}]_{\alpha \beta}^\dagger c_{i \alpha}, \quad \bar c_{j \beta} = [\hat U^{ij}]_{\alpha \beta}c_{j \beta}, \quad \hat U_{ij} = e^{i\phi_{ij} \boldsymbol u_{ij} \cdot \boldsymbol \sigma/2}.
\end{align}
In terms of the rotated operators, $H$ becomes
\begin{eqnarray}
\hat H &=& -\bar t \sum_{\langle i, j \rangle, \alpha} \bar c^\dagger_{i\alpha} \bar c_{j \alpha} + \sum_{\langle i, j \rangle} J\bar {\boldsymbol S}_i \cdot \bar {\boldsymbol S}_j - 2 J_H \bar{\boldsymbol s}_i \cdot \bar{\boldsymbol S}_j \delta _{ij} \nonumber \\
&= &-\bar t \sum_{\langle i, j \rangle, \alpha} \bar c^\dagger_{i\alpha} \bar c_{j \alpha} + \sum_{\langle i,j \rangle} J(\hat R_{ij}^T{\boldsymbol S_i}) \cdot (\hat R_{ij}{\boldsymbol S_j)} \nonumber \\ 
&&-2 J_H(\hat R_{ij}^T{\boldsymbol s_i}) \cdot (\hat R_{ij}{\boldsymbol S_j)} \delta_{ij},\label{eq:And}
\end{eqnarray}
where $\bar t = \sqrt{t^2 + \lambda_\text{SO}^2}$ and $\hat R_{ij} \in \text{SO(3)}$ represents the same rotation as $\hat U_{ij}$ in the three-dimensional Euclidean space: a rotation by $-\phi_{ij}$ around the $\boldsymbol u_{ij}$ axis, which can be written as
\begin{align}
\hat R_{ij} = \cos\phi_{ij} \openone + (1 - \cos\phi_{ij})\boldsymbol u_{ij} \boldsymbol u_{ij}^T - \sin\phi_{ij} \boldsymbol u_{ij} \times
\label{eq:so3expansion}
\end{align}
In the limit of large $J_H$ and classical spins $\bar{\boldsymbol S_i}$,  Hamiltonian  (\ref{eq:And}) corresponds to effective exchange interaction between localized spins given by \cite{Anderson1955}
\begin{align}
-J_F \sum_{\langle i, j \rangle}  \sqrt{1 + \bar{\boldsymbol S}_i \cdot \bar{\boldsymbol S}_j/2 S^2},
\end{align}
where $S$ is the magnitude of the local spins, $J_F = \mathcal K \bar t$, and $\mathcal K$ is a constant related to the density of itinerant electrons.
Expanding the square root and using $(\hat R_{ij}^T{\boldsymbol S_i}) \cdot (\hat R_{ij}{\boldsymbol S_j)} = {\boldsymbol S_i} \cdot (\hat R^2_{ij}{\boldsymbol S_j)}$ along with Eq.~(\ref{eq:so3expansion}) and considering nearest-neighbor interactions, we obtain the Hamiltonian
\begin{align}
\hat H = \sum_{i,\mu = x, y} & -J^\mu \boldsymbol S_i \cdot \boldsymbol S_{i+\boldsymbol \mu} 
-D^\mu  \boldsymbol u_\mu \cdot  (\boldsymbol S_i \times \boldsymbol S_{i+\boldsymbol \mu}) \nonumber \\
&- A_c^\mu ( \boldsymbol u_\mu\cdot \boldsymbol S_i) (\boldsymbol u_\mu \cdot \boldsymbol S_{i+\boldsymbol \mu})
\end{align}
with $J^\mu = -J \cos2\phi_\mu$, $A_c^\mu = J_o(1-\cos2\phi_\mu)$, and $D^\mu = J \sin2\phi_\mu$.
Since $\phi_\mu \sim \lambda_\text{SO}/t \ll 1$, we obtain the relation $A_c J/D^2  \approx 1/2$, where $A_c \approx A_c^\mu (D/|\hat D \boldsymbol e_\mu|)^2 $.

The continuous free energy density corresponding to this lattice tight-binding model is
\begin{align}
\mathcal F = &  
 \sum_{ \mu =x, y} \frac{J}{2}(\partial_\mu \boldsymbol n)^2 
 + (\hat D \boldsymbol e_\mu) \cdot(\boldsymbol n \times \partial_\mu \boldsymbol n) \nonumber\\
&- A_c \left[ \left(\boldsymbol {\mathfrak D}_\mu \cdot \boldsymbol n\right)^2 
- \frac{1}{2}\left(\boldsymbol {\mathfrak D}_\mu \cdot \partial_\mu \boldsymbol n\right)^2  - \frac{1}{2}\left(|\boldsymbol {\mathfrak D}_\mu |\partial_\mu \boldsymbol n\right)^2 \right].
\end{align}
Using $J/D$ as the new unit of length, the free energy density can be rewritten in the dimensionless form
\begin{align}
\mathcal F_0 = &\sum_{ \mu = x, y} \frac{1}{2}(\partial_\mu \boldsymbol n)^2 
 + \boldsymbol {\mathfrak D}_\mu \cdot(\boldsymbol n \times \partial_\mu \boldsymbol n) - \frac{A_cJ}{D^2} \left(\boldsymbol {\mathfrak D}_\mu \cdot \boldsymbol n\right)^2
 \nonumber\\
& +\frac{A_cJ}{D^2} 
 \frac{1}{2}\left(\frac{D}{J}\right)^2 \left[(\boldsymbol {\mathfrak D}_\mu \cdot \partial_\mu \boldsymbol n)^2 + \left(|\boldsymbol {\mathfrak D}_\mu | \partial_\mu \boldsymbol n\right)^2 \right] 
\label{eq:microscopic-F}
\end{align}
in units of $J$. After adding the uniaxial anisotropy and Zeeman energies as well as dropping the  term $(D/J)^2 \sim (\lambda_\text{SO}/t)^2 \ll 1$, this free energy density shares the symmetries of the one given by Eq.~(\ref{eq:free-energy}). They become equivalent for $\hat D = -J_z$ or $\hat D = -\openone$ leading to a uniaxial anisotropy with a renormalized strength $A_s \to A_s + A_c$. In the case of $C_{2v}$ symmetry discussed in Sec.~\ref{sec:phase-diagrams}, there is an additional anisotropy term $A_c 2 D_R D_D (D_R^2 + D_D^2)^{-1} (n_x ^2 - n_y^2)$ which is compatible with $C_{2v}$ symmetry. For the case with only reflection symmetry, the leading anisotropy term is $A_c 2\theta_T n_z (n_x \cos \phi_T + n_y \sin \phi_T) $, where $\theta_T$ and $\phi_T$ are spherical angles describing the tilting vector $\boldsymbol n_T$; as one would expect, the additional term respects the reflection symmetry. We numerically found that for small $D_D/D_R$ or $\theta_T$ there is no significant difference in the phase diagrams due to additional anisotropy terms.

\section{Variational estimate}
In order to obtain an initial rough estimate for the phase diagram, we minimize the average free energy density
\begin{align}
\frac{F}{A} =\frac{1}{A}\int_A d^2\boldsymbol r \mathcal F(\boldsymbol n(\boldsymbol r))
\end{align}
by comparing the energies corresponding to ferromagnetic, SkX, SC, and SP ansatzs. For each ansatz we optimize the parameters that yield the minimal energy.

\emph{Skyrmion lattice.} We assume a simple form of azimuthally symmetric skyrmion and linearly interpolate the $n_\theta$ component of the spin density:
\begin{align}
\boldsymbol n_\text{SkX} = (\sin n_\theta \cos n_\phi, \sin n_\theta \sin n_\phi, \cos n_\theta)^T
\end{align}
with $n_\theta = -\pi r/R$ and $n_\phi = Q \phi + \gamma$. The optimal values of $R$ (skyrmion size), $Q \in \{-1,1\}$ (topological charge) and $\gamma$ (helicity) are found by minimization. In the reduced symmetry cases, the accuracy of this azimuthally symmetric ansatz becomes less accurate as the asymmetry is increased. In principle, this can be remedied by adding a constant component to $\boldsymbol n_\text{SkX}$ along a preferred direction.

\emph{Spiral.} MC simulations show that depending on $\hat D$, spirals can be coplanar or in-plane. To capture both kinds of spirals, we assume the ansatz
\begin{align}
\boldsymbol n_\text{SP} = \boldsymbol e_u \cos(\boldsymbol q \cdot \boldsymbol r) + \boldsymbol e_v \sin(\boldsymbol q \cdot \boldsymbol r)
\end{align}
for the spiral phase with $\boldsymbol e_u \cdot \boldsymbol e_v = 0$. Minimization is done with respect to $\boldsymbol e_u$, $\boldsymbol e_v$, and $\boldsymbol q$.

\emph{Ferromagnetic.} There are two types of uniform ferromagnetic phase that appear as the solution of equation:
\begin{align}
A_s n_z^2 - H n_z = 0.
\end{align}
The first case is the aligned ferromagnetic phase with $n_z=-1$ with energy $A_s+H$. In the tilted ferromagnetic phase, we have $n_z = -H/2 A_s > -1$ and the energy is $-H^2/4A_s$.
The resulting phase diagrams coarsely agree with the results obtained from using LLG equation and MC.

\section{LLG equation with magnonic torques}
\label{sec:LLG}
In this Appendix, we give a derivation of the LLG equation (\ref{eq:LLG}) when the spin current and torque originate from magnon currents. The presentation here follows closely \cite{Kovalev2014a,Kovalev2014b,Kovalev2015a} and extends the result in \cite{Kim2015} to systems with DM interaction.

We assume that the time scale of the magnetization dynamics determined by external magnetic fields, anisotropies, and currents is slow compared to the time scale of thermal magnons defined by the temperature. We also assume that the thermal magnon wavelength is smaller than the typical size of the magnetic texture, i.e., skyrmion.
Magnetization dynamics of a ferromagnet well below Curie temperature can
 be described by the stochastic LLG equation
\begin{align}
s(1+\alpha \boldsymbol n \times) \dot {\boldsymbol n} =  \boldsymbol n \times (\boldsymbol H_\text{eff} + \boldsymbol h) ,
\label{eq:sLLG}
\end{align}
where $s$ is the spin density, $\boldsymbol n = \boldsymbol n(\boldsymbol r, t)$ is a unit vector along spin density, $\boldsymbol H_\text{eff} = -\delta_{\boldsymbol n} F$ is the effective magnetic field and $\boldsymbol h$ is the random Langevin field at temperature $T$ that satisfies the fluctuation-dissipation theorem with the correlator
\begin{align}
\langle h_i(\boldsymbol r, t) h_j(\boldsymbol r, t) \rangle = 2 \alpha s T(\boldsymbol r) \delta_{ij} \delta(\boldsymbol r - \boldsymbol r') \delta(t - t').
\end{align}
Here $\alpha$ is the Gilbert damping coefficient and we assume a uniform temperature gradient along the $x$-axis, $\partial_x T = \text{const}$.

We separate the spin density $\boldsymbol n$ into small and fast oscillations $\boldsymbol n_s $ with time scale $1/\omega_k$ ($\omega_k$ is the magnon frequency) on top of the slow spin density dynamics $\boldsymbol n_f$ whose time scale is defined by external magnetic fields, anisotropies, and currents. These two orthogonal (that is, $\boldsymbol n_f \cdot \boldsymbol n_s = 0$) components are related to spin density by $\boldsymbol n =  \sqrt{1-\boldsymbol n_f^2} \boldsymbol n_s + \boldsymbol n_f$ by definition, and $|\boldsymbol n_s| = |\boldsymbol n|=1$.
We now switch to a coordinate system where the $z$-axis points along $\boldsymbol n_s$ through a local SO(3) gauge transformation $\hat R_s = \exp(\Psi \hat J_z) \exp(\Theta \hat J_y) \exp(\Phi \hat J_z)$, where $\Psi, \Theta$, and $\Phi$ are Euler angles. In what follows we set $\Psi= 0$. In this coordinate system, the spin density becomes $\boldsymbol n' = \hat R_s \boldsymbol n$ and the covariant derivative is $\partial_\nu - \hat {\mathcal A}_\nu$ with $\hat {\mathcal A}_\nu = (\partial_\nu \hat R_s) \hat R_s^T = \boldsymbol {\mathcal A}_\nu \times$ and $\boldsymbol {\mathcal A}_\nu = (-\sin\Theta \partial_\nu \Phi, \partial_\nu \Theta, \cos\Theta \partial_\nu \Phi)$, and $\nu = t,x,y,z$ denotes time and space coordinates. By treating $\boldsymbol {\mathcal A}_\nu$ as a (fictitious) vector potential, we fix the gauge.
In the new frame $\boldsymbol n_f$ lies in the $x'-y'$ plane which can be represented by a complex number as $n_f = n_{x'} + i n_{y'}$.
The LLG equation in this frame for the fast dynamics is given by
\begin{align}
i s [(1- i \alpha)\partial_t - i \mathcal A_t^z]n_+ =& J(i \partial_\mu + \mathcal A_\mu ^ z - d_\mu^z)^2 n_+  + H n_+,
\label{eq:LLG-fast-dynamics}
\end{align}
where $\boldsymbol d_\mu = \hat D \boldsymbol e_\mu/J$ and we assumed exchange interactions are dominant over various anisotropy terms and the coupling between the circular components $n_f$  and $n_f^*$ can be ignored \cite{Dugaev2005,Kovalev2014a}. RHS of this equation can be read as follows: the gauge fields $\boldsymbol {-\mathcal A}_\mu$ and $\hat D \boldsymbol e_\mu/J$, respectively, account for aligning $\boldsymbol n_s'$ with the $z'$-axis and introducing the DM interaction; the gauge potentials can be merged into an overall covariant derivative 
$\mathcal D_\mu' \approx \partial_\mu + (- \boldsymbol {\mathcal A}_\mu + \hat D \boldsymbol e_\mu/J) \times$,
where we kept only the first order terms in $D/J$, $\partial_\mu \Theta$ and $\partial_\mu \Phi$. Equation~(\ref{eq:LLG-fast-dynamics}) describes thermal magnons with spectrum $\omega_k = [H + J (\boldsymbol k-\boldsymbol k_0)^2 - J k_0^2]/s$ and magnon current $j_\mu = J\, \text{Im} (n_f^* \partial_\mu n_f)$. Here, $\boldsymbol k_0$ is the momentum shift induced by the DM interaction and magnetic texture. Formally, Eq.~(\ref{eq:LLG-fast-dynamics}) describes the motion of charged particles due to fictitious electric and magnetic fields $\mathcal E_\mu = \boldsymbol n'_s \cdot (\partial_t \boldsymbol n'_s \times \partial_\mu \boldsymbol n'_s)$ and $\mathcal B_i = -(\epsilon_{ijk}/2) \boldsymbol n'_s \cdot (\partial_j \boldsymbol n'_s \times \partial_k \boldsymbol n'_s)$.

We are now in a position to calculate the force exerted by the fast oscillations on the slow spin density dynamics. For simplicity, we assume that the slow magnetic texture is static since the time-dependence can be taken into account later on by Onsager reciprocity principle. The relevant terms in the total effective field
\begin{align}
\boldsymbol H_\text{eff} = J \left( \partial_\mu^2 \boldsymbol n + 2 \frac{\hat D \boldsymbol e_\mu}{J} \times\partial_\mu \boldsymbol n \right) - (H + 2 A_s n_z)\boldsymbol e_z
\end{align}
are exchange and DM terms because other terms  average out due to rapid oscillations (summation over $\mu=x,y$ is implied), thus we obtain the expression for the torque
\begin{align}
\boldsymbol {\mathcal T} &=   \langle \boldsymbol n \times \boldsymbol H_\text{eff} \rangle - \langle n \rangle \boldsymbol n_s \times \boldsymbol H_\text{eff}^s =
J\boldsymbol n_s \times \boldsymbol {\mathcal S},  \nonumber \\
&\approx J\langle \boldsymbol n_f \times \partial_\mu^2 \boldsymbol n_f\rangle + 2 J \langle \boldsymbol n \times (\partial_\mu \boldsymbol n_s) \partial_\mu (\boldsymbol n_s \cdot \boldsymbol n) \rangle  \nonumber \\
& +2 J \langle \boldsymbol n_f \times (\boldsymbol d_\mu \times \partial_\mu \boldsymbol n_f) \rangle  ,
\end{align}
where we formally introduced $\boldsymbol {\mathcal S} = -\boldsymbol n_s \times \boldsymbol {\mathcal T}/J$ as the nonequilibrium transverse accumulation of magnon spins, $\langle \ldots \rangle$ denotes averaging over the fast oscillations induced by random fields and $\boldsymbol H_\text{eff}^s = -\delta_{\boldsymbol n_s} F(\langle n \rangle \boldsymbol n_s, \langle n \rangle \partial_\mu \boldsymbol n_s)$. By dropping oscillatory terms that average out, we obtain
\begin{align}
\boldsymbol {\mathcal S} =& 2 \langle \boldsymbol n_f (\partial_\mu \boldsymbol n_s \cdot \partial_\mu \boldsymbol n_f) \rangle - 2 \langle \boldsymbol n_s (\partial_\mu \boldsymbol n_f \cdot \partial_\mu \boldsymbol n_f)\rangle  \nonumber \\
&+2 \langle(\boldsymbol n_s \times \boldsymbol n_f)(\boldsymbol d_\mu \cdot \partial_\mu \boldsymbol n_f)\rangle.
\label{eq:Svec}
\end{align}
The fact that vectors $\boldsymbol {\mathcal S}, \boldsymbol  n_f, \partial_\mu \boldsymbol  n_s, \partial_\mu \boldsymbol  n_f$ and $\boldsymbol n_s \times \boldsymbol n_f$ are in the $x'-y'$ plane allows us to rewrite Eq.~(\ref{eq:Svec}) using complex numbers, leading to
\begin{align}
\mathcal S = -i d_\mu \langle n_+ \partial_\mu n_- \rangle - (\partial_\mu n_s)\langle n_- \partial_\mu n_+ \rangle.
\end{align}
$\mathcal S = \mathcal S_x + i \mathcal S_y$ describes the components of spin accumulation leading to dissipative and nondissipative torques and $-i d_\mu$ represents $\boldsymbol d_\mu \times \boldsymbol n_s$ (which also is a vector in the $x'-y'$ plane) as a complex number. For the steady state solution, we obtain
\begin{align}
\langle n_\pm \partial_\mu n_\mp \rangle = & \pm \int \frac{d^{d-1} \boldsymbol k d \omega}{(2\pi)^d}\frac{\langle n_\pm(\boldsymbol k, \omega, x) \partial_\mu n_\mp(\boldsymbol k', \omega', x)\rangle}{(2\pi)^d \delta(\boldsymbol k-\boldsymbol k') \delta(\omega - \omega') },
\label{eq:S}
\end{align}
where $d=2,3$ is the dimensionality of the magnet and
\begin{align}
n_\pm(\boldsymbol k, \omega, x) = \int \frac{d^{d-1} \boldsymbol \rho d \omega}{(2\pi)^d} e^{\mp i (\omega t - \boldsymbol \rho \cdot \boldsymbol k)} n_\pm(\boldsymbol \rho, t)
\end{align}
is the Fourier transform with respect to time and transverse coordinates.
The $\delta$-functions in the denominator are canceled by the stochastic field
\begin{align}
\frac{\langle h^*(\boldsymbol k, \omega, x) h(\boldsymbol k', \omega', x) \rangle}{4 (2 \pi)^d \alpha s k_B} = T(x) \delta(x-x') \delta (\omega-\omega').
\end{align}
Since we are interested in the linear response to the random Langevin field $h(\boldsymbol k, \omega, x)$, we set $\mathcal A_\nu^z=0$ in Eq.~(\ref{eq:LLG-fast-dynamics}) and the stochastic LLG equation Eq.~(\ref{eq:sLLG}) becomes the inhomogeneous Helmholtz equation
\begin{align}
J(\partial_x^2 + \kappa^2)n_-(\boldsymbol k, \omega, x) = h(\boldsymbol k, \omega, x),
\end{align}
where $\kappa^2 = [(1 + i \alpha) s \omega-H]/J-k^2-k_0^2$. This equation corresponds to Eq.~(\ref{eq:LLG-fast-dynamics}) with an added stochastic term, and can be solved easily by employing Green's function $G(x-x_0) = i e^{i \kappa |x - x_0|}/2\kappa$. By substituting the solution into Eq.~(\ref{eq:S}) and employing the quantum dissipation theorem, we find
\begin{align}
\mathcal T = - (\mathcal D_\mu n_s) j_x (1 + i \beta),
\label{eq:T}
\end{align}
where $\mathcal D_\mu n_s$ represents $\mathcal D_\mu \boldsymbol n_s$ as a complex number and
\begin{align}
j_x = \frac{\partial_x T}{T} \int \frac{d^d \boldsymbol k}{(2\pi)^d} \tau(\epsilon) \epsilon v_x^2 \partial_\epsilon f_0
\end{align}
with $\tau(\epsilon) = (2\alpha \omega)^{-1}$, $\epsilon(\boldsymbol k) = (J k^2 + H)/s$, $v_x = \partial \omega_k / \partial k_x$ and $f_0 = 1/[\exp(\epsilon/k_B T) - 1]$ is the Bose-Einstein equilibrium distribution. The $\beta$ term in Eq.~(\ref{eq:T}) corresponds to the dissipative correction with  $\beta/\alpha= (d/2) F_1(x)/F_0(x) \sim d/2$ with $F_0(x) = \int_0^\infty d \epsilon \epsilon^{d/2-1} \epsilon e^{\epsilon+x}/ (e^{\epsilon+x}-1)^2$ and $F_1(x) = \int_0^\infty d \epsilon \epsilon^{d/2-1} (\epsilon+x) e^{\epsilon+x}/ (e^{\epsilon+x}-1)^2$ evaluated at the magnon gap $x = \omega_0 / k_B T$ (for $d>2$ and for small gaps, $F_1(x) = F_0(x) = \zeta(d/2)/\Gamma(1+d/2)$, where $\zeta(x)$ is the Riemann zeta function, and $\Gamma(x)$ is the Euler gamma function \cite{Pathria1996}). The magnon current density is given by
\begin{align}
j_\mu = -k_B (\partial_\mu T) F_0 / (6 \pi^2 \lambda \hbar \alpha)
\end{align}
for $d=3$, where $\lambda = \sqrt{\hbar J/s k_B T}$ is the thermal magnon wavelength, and
\begin{align}
j_\mu = -k_B (\partial_\mu T) F_0 /( 4 \pi \hbar \alpha)
\end{align}
for $d=2$. Using the spin-torque term given by Eq.~(\ref{eq:T}), we obtain the LLG equation with thermomagnonic torque
\begin{align}
\mathfrak s (1+ \alpha^s \boldsymbol n_s \times) \dot {\boldsymbol n_s}  =  \boldsymbol n_s \times \boldsymbol H_\text{eff}^s &- (1 +\beta \boldsymbol n_s \times) (\boldsymbol j^s \cdot \boldsymbol {\mathcal D} ) \boldsymbol n_s,
\end{align}
where $\mathfrak s = \langle n \rangle s$ is the renormalized spin density, $\boldsymbol H_\text{eff}^s = -\delta_{\boldsymbol n_s} F(\langle n \rangle \boldsymbol n_s, \langle n \rangle \partial_\mu \boldsymbol n_s)$ is the effective field, $\alpha^s = \langle  n \rangle \alpha$ is the renormalized Gilbert damping, and $\boldsymbol j^s = -\hbar \boldsymbol j$ is the spin current with polarization along $\boldsymbol n_s$ carried by magnons.

\section{Scaling of the lattice vectors}
\label{sec:scaling}
In order to determine the ground state of the system at a given point $(A,H)$ in the phase diagram, we compare the energies of relaxed SP, SC, SkX, and FM configurations.
We use LLG equation to simulate the spin dynamics and to relax the system starting from ansatz-states (SP, SC and SkX as given by Eq.~(\ref{eq:ansatz}), which are obtained by inspecting the results from MC). We then compare the resulting average free energy densities and determine the actual phase.

In each case, we relax a single primitive cell with rectangular periodic boundary conditions. This, however, requires us to specify the cell size, which we do not know beforehand. To overcome this problem, we dynamically scale the spatial coordinates as $x' = a_x x$ and $y' = a_y y$ such that the lattice vectors are scaled to their optimal values as we describe below.

The dimensions of the primitive cell $(L_x,L_y)$ in the $(x',y')$ coordinate system are chosen such that the ansatz (SkX, SC, SP) is compatible with periodic boundaries (that is, $L_x=\pi$, $L_y=\pi/\sqrt{3}$ for SkX, $L_x=L_y=\pi/2$ for SC and $L_x=L_y=\pi$ for SP).
The average free energy density is given by $f=F/A$, where
\begin{align}
F =& J\int_{-L_y}^{L_y} \int_{-L_x}^{L_x} dx' dy' |\mathcal J| \mathcal F', \nonumber \\
A =& \int_{-L_y}^{L_y} \int_{-L_x}^{L_x} dx' dy' |\mathcal J|, \nonumber \\
\mathcal F' =& \sum_{\mu = x,y} \frac{1}{2} (a_\mu \partial_{\mu'} \boldsymbol n)^2 + (\hat D \boldsymbol e_\mu/D) \cdot (\boldsymbol n \times a_\mu\partial_{\mu'} \boldsymbol n)  \nonumber \\
& + \frac{H J}{D^2} n_z + \frac{A_s J}{D^2} n_z^2, \nonumber \\
|\mathcal J| =& \frac{1}{a_x a_y}.
\end{align}
Here $|\mathcal J|$ is the Jacobian of the transformation. The optimal value of $a_\mu$ is given by $\partial_{a_\mu} f= 0$ or
\begin{align}
a_\mu = -\frac{1}{2}
\frac{\int_{-L_y}^{L_y} \int_{-L_x}^{L_x} dx' dy' |\mathcal J| (\hat D \boldsymbol e_\mu/D) \cdot (\boldsymbol n \times \partial_{\mu'} \boldsymbol n)}
{\int_{-L_y}^{L_y} \int_{-L_x}^{L_x} dx' dy' |\mathcal J|\frac{1}{2}(\partial_{\mu'} \boldsymbol n)^2}.
\label{eq:mina}
\end{align}

Starting from an ansatz configuration $\boldsymbol n^{(0)}$ given by Eq.~(\ref{eq:ansatz}), one can determine the optimal value of $a_\mu^{(0)}$ using Eq.~(\ref{eq:mina}) and relax the system using the overdamped LLG equation Eq.~(\ref{eq:overdamped-LLG}) for a period of time $T_R$. This process yields the configuration $\boldsymbol n^{(1)}$ which is closer to the local minimum, and we can calculate $a_\mu^{(1)}$ using $\boldsymbol n^{(1)}$. Eventually, $| f[\boldsymbol n^{(i+1)}, a_\mu^{i+1}] - f[\boldsymbol n^{(i)}, a_\mu^{i}]|$ vanishes as we approach the minimum.

A corollary of Eq.~(\ref{eq:mina}) is that at the equilibrium point, 
\begin{align}
\frac{F_D^\mu}{2 F_J^\mu} = \frac{\int_{-L_y}^{L_y} \int_{-L_x}^{L_x} dx' dy' |\mathcal J| (\hat D \boldsymbol e_\mu/D) \cdot (\boldsymbol n \times a_\mu \partial_{\mu'} \boldsymbol n)}
{2 \int_{-L_y}^{L_y} \int_{-L_x}^{L_x} dx' dy' |\mathcal J| \frac{1}{2}(a_\mu \partial_{\mu'} \boldsymbol n)^2} = -1,
\end{align}
and since $[\hat \eta]_{\mu\mu} = 2 F_J^\mu$ and $[\hat \eta_1]_{\mu\mu} = 2 F_J^\mu + F_D^\mu$, we reach to the conclusion that $[\hat \eta_1]_{\mu\mu}=0$.

We emphasize that this result is derived under the assumption of a periodic lattice and does not hold for isolated skyrmions in general.

\vfill
\bibliographystyle{apsrev}
\bibliography{skyrmion}
\end{document}